\documentclass{article}

\usepackage[preprint, nonatbib]{neurips_2021}

\usepackage[utf8]{inputenc} 
\usepackage[T1]{fontenc}    
\usepackage{hyperref}       
\usepackage{url}            
\usepackage{booktabs}       
\usepackage{amsfonts}       
\usepackage{nicefrac}       
\usepackage{microtype}      
\usepackage{xcolor}         

\usepackage{tikz}
\usetikzlibrary{calc}
\usepackage{latexsym}
\usepackage{graphicx}
\usepackage{multicol,multirow}
\usepackage{amsmath,amssymb,amsfonts}
\usepackage{mathrsfs}
\usepackage{amsthm}
\usepackage{rotating}
\usepackage{appendix}
\usepackage{ifpdf}
\usepackage[T1]{fontenc}
\usepackage{times}
\usepackage{newtxmath}
\usepackage{textcomp}%
\usepackage{xcolor}%
\usepackage{hyperref}
\usepackage{lipsum}

\usepackage[authoryear]{natbib}
\setcitestyle{authoryear,open={(},close={)}}
\usepackage{amsmath}
\usepackage{amssymb}

\usepackage{graphicx}
\usepackage{parskip}

\usepackage{booktabs}

\usepackage{algorithm}
\usepackage{algpseudocode}

\DeclareMathOperator{\sign}{sign}

\newcommand\var{\mathrm}

\newcommand{\R}{\mathbb{R}}

\title{AtmoDist: Self-supervised Representation Learning for Atmospheric Dynamics}

\author{%
  Sebastian Hoffmann \\
  Dept. of Computer Science\\
  Universit{\"a}t Magdeburg\\
  \texttt{sebastian1.hoffmann@ovgu.de} \\
 \And
 Christian Lessig \\
  Dept. of Computer Science\\
  Universit{\"a}t Magdeburg\\
  \texttt{christian.lessig@ovgu.de}
}

\begin{document}

\maketitle

\begin{abstract}
Representation learning has proven to be a powerful methodology in a wide variety of machine learning applications.
For atmospheric dynamics, however, it has so far not been considered, arguably due to the lack of large-scale, labeled datasets that could be used for training. 
In this work, we show that the difficulty is benign and introduce a self-supervised learning task that defines a categorial loss for a wide variety of unlabeled atmospheric datasets.
Specifically, we train a neural network on the simple yet intricate task of predicting the temporal distance between atmospheric fields from distinct but nearby times. 
We demonstrate that training with this task on ERA5 reanalysis leads to internal representations capturing intrinsic aspects of atmospheric dynamics.
We do so by introducing a data-driven distance metric for atmospheric states. 
When employed as a loss function in other machine learning applications, this Atmodist distance leads to improved results compared to the classical $\ell_2$-loss.
For example, for downscaling one obtains higher resolution fields that match the true statistics more closely than previous approaches and for the interpolation of missing or occluded data the AtmoDist distance leads to results that contain more realistic fine scale features.
Since it is derived from observational data, AtmoDist also provides a novel perspective on atmospheric predictability. 
\end{abstract}

\section{Introduction}

Representation learning is an important methodology in machine learning where the focus is on the data transformations that are provided by a neural network. 
The motivation for it is to obtain an embedding of the input data that will facilitate a range of applications, e.g. by revealing intrinsic aspects of it or by being invariant to perturbations that are irrelevant for tasks.
Representation learning is today central to application areas such as machine translation, e.g.~\cite{Devlin2019}, and image understanding, e.g.~\cite{Caron2021,bao2022beit}, and has led there to significantly improved performance on a variety of tasks. 

In the geosciences, representation learning has so far received only limited attention.
One reason is the lack of large-scale, labeled data sets that are classically used for training. 
As has been shown for other domains, e.g.~\cite{he2020momentum,Caron2021}, representation learning can, however, benefit from working with unlabeled data and performing self-supervised learning with a loss functions derived from the data itself.
One reason for this is that a self-supervised task can be more challenging than, e.g., choosing from a small set of possible answers or labels.
Hence, with a self-supervised task the neural network is forced to learn more expressive and explanatory internal representations.
A second reason for the efficiency of self-supervised training is that it makes much larger amounts of data available since no labels are required, e.g.~\cite{Devlin2019,Zhai2021}.

\begin{figure}
  \centering
  \includegraphics[width=1.\textwidth]{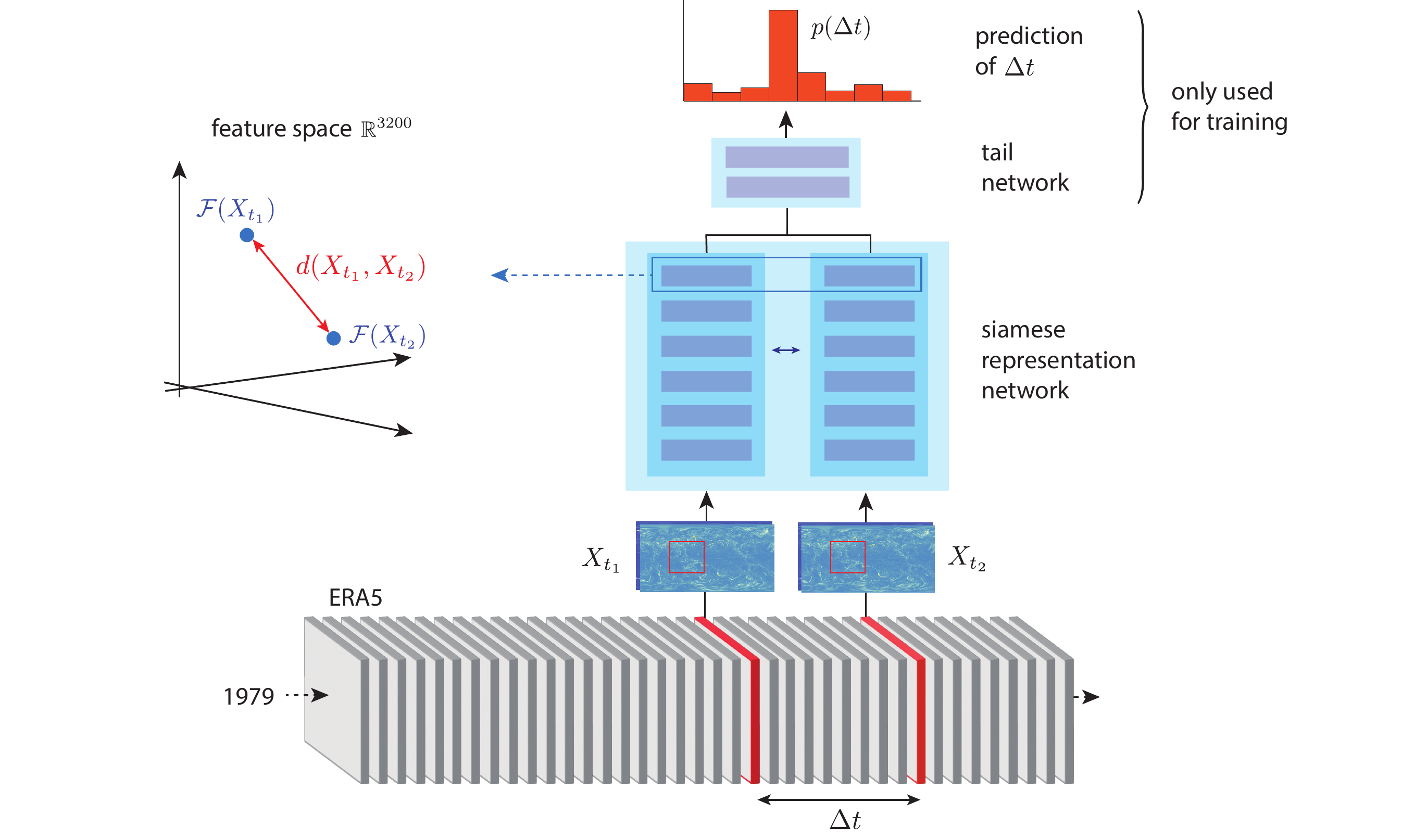}
  \caption{Overview of the methodology for AtmoDist. From a temporal sequence of atmospheric fields (bottom), two nearby ones are selected at random (red) and stored together with their temporal separation $\Delta t$ as a training sample. Both fields are then passed through the same representation network (blue), which embeds them into a high-dimensional feature space (left). These embeddings are subsequently used by the tail network to predict the temporal separation $\Delta t$ (top, orange). The whole architecture is trained end-to-end. Once training is done, the embeddings can be used in downstream tasks, e.g. through a distance measure $d(X_{t_1}, X_{t_2})$ in embedding space} 
  \label{fig:atmodist}
\end{figure}

In this work, we introduce self-supervised representation learning for atmospheric dynamics and demonstrate its utility by defining a novel, data-driven distance metric for atmospheric states based on it.
For the self-supervised training, we propose a novel learning task that is applicable to a wide range of data sets in atmospheric science.
Specifically, given a temporal sequence of datums, e.g. spatial fields in a reanalysis or from a simulation, the task of the neural network is to predict the temporal distance between two randomly selected, close-by sequence elements. 
Performing well on the task requires the network to develop an internal representation of the underlying dynamics, which will typically be useful for a variety of tasks. 

We demonstrate the effectiveness and practicality of our self-supervised training task by learning a representation network for vorticity and divergence (which are equivalent to the wind velocity field) from ERA5 reanalysis~\citep{era5}. 
From the learned representation, we subsequently derive a data-driven distance metric for atmospheric states, which we call the AtmoDist distance.
To demonstrate its potential, we use it as loss function in GAN-based downscaling as well as to interpolate missing data. 
Building on the state-of-the-art GAN by~\cite{Stengel2020} we show that the AtmoDist loss significantly improves downscaling results compared to the $\ell_2$ loss used in the original work.
For missing data interpolation, AtmoDist leads to a more realistic fine scale details and better local statistics.
As baseline we use in these experiments an auto-encoder, which provides an alternative means to obtain a feature space representation using self-supervised training.
We also report results on experiments with AtmoDist on the predictability of atmospheric states where the data-driven loss reproduces known dependencies on season and spatial location.

We believe that self-supervised representation learning for atmospheric data has significant potential and we hence consider the present work as a first step in this direction. 
Self-supervised learning only requires unlabeled data, which is available in significant quantities, e.g. in the form of satellite observations, reanalyses and simulation outputs.
Given the difficulty of obtaining large, labeled data sets, this removes an obstacle for the use of large scale machine learning in the atmospheric sciences. 
At the same time, representation learning can ``distill'' effective representations from very large amounts of data~\citep{Devlin2019,Zhai2021}, which might, e.g., provide a new avenue to process the outputs produced by large simulation runs~\citep{Eyring2016}.
We believe that learned representation can also be useful to gain novel scientific insights into the physics, similar to how proper orthogonal decompositions have been used in the past.
This is, in our opinion, a particularly inspiring direction for future work~\citep{Toms2020}.

\section{Related Work}
\label{sec:related}

In the following, we will discuss pertinent related work from both geoscience and machine learning.

\subsection{Geoscience}

Distance measures for atmospheric states play an important role in classical weather and climate predictions.
For example, ensemble methods require a well defined notion of nearby atmospheric states for their initialization.
Various distance measures have therefore been proposed in the literature, typically grounded in mathematical and physical considerations, e.g. conservation laws.
The importance of an appropriate distance measure for atmospheric states already appears in the classical work by~\cite{Lorenz1969} where atmospheric predictability depends on the closeness of initial states and is also affected by the characteristics of their spectrum, i.e. a Sobolev-type measure.
\cite{Talagrand1981} considered an energy metric around a reference state obtained from the primitive equations in work on 4D data assimilation. 
Palmer and co-workers~\citeyearpar{Palmer1998} argue that within the framework of linearized equations and with singular vectors as coordinates, a metric for targeting observations should not only be informed by geophysical fluid dynamics considerations but also consider the operational observing network.
Recently,~\cite{Koh2015} introduce an energy metric that does not require an atmospheric reference state but is intrinsically defined.
For the case of an ideal barotropic fluid, the metric of~\cite{Koh2015} also coincides with the geodesic metric that was introduced by~\cite{Arnold1966} and studied by~\cite{Ebin1970} to describe the fluid motion as a geodesic on the infinite dimensional group of volume preserving diffeomorphisms.
Although of utility in classical applications, the aforementioned distance measures  lack the sensitivity desirable for machine learning techniques, e.g. with respect to small scale features, and are agnostic to applications.
In the context of downscaling, this deficiency has recently been noted by~\cite{Stengel2020}.

\subsection{Representation Learning and Learned Distance Measures}

Representation learning~\citep{Bengio2013} focuses on the nonlinear transformations that are realized by a neural network and understands these as a mapping of the input data to a feature space adapted to the data domain. 
The feature space is informative and explanatory, for example, when different classes are well separated and interdependencies are transparently encoded.
This then allows to solve so-called downstream applications in a simple and efficient manner, for example by appending a linear transformation or a very small neural network to the pre-trained one.
Good representations will also be useful for a wide range of applications.

A pertinent example for the important role of representations in neural networks is classification.
There, the bulk of the overall network architecture is usually devoted to transforming the data into a feature space where the different classes correspond to linear and well-separated subspaces.
A linear mapping in the classification head then suffices to accurately solve the task and the entire preceding network can thus be considered as a representational one. 
With deep neural networks, one obtains a hierarchy of representations where deeper once typically correspond to more abstract features, see e.g.~\cite{Zeiler2014} for visualizations.
The hierarchical structure is of importance for example for generative machine learning models, e.g.~\citep{Ronneberger2015,Karras2019,Karras2020,Ranftl2021} where features at all scales have to match the target distribution.

An important applications of representation learning is the design of domain-specific loss functions, sometimes also denoted as content losses~\citep{zhang2018unreasonable}. 
The rationale for these is that feature spaces are designed to capture the essential aspects of an input data domain and computing a distance there is hence more discriminative than on the raw inputs~\citep{achille2018emergence}. 
Furthermore, deeper layers typically have invariance against ``irrelevant'' perturbations, such as translation, rotation, and noise.
A classical example are natural images where $\ell_p$-norms in the pixel domain are usually not well suited for machine learning, e.g. because a small shift in the image content can lead to a large distance in an $\ell_p$-norm despite the image being semantically unchanged.
Loss functions computed in the feature spaces of networks such as VGG~\citep{Simonyan2015}, in contrast, can lead to substantially improved performance in task such as in-painting~\citep{Yang2017a}, style transfer~\citep{Gatys2016}, and image synthesis~\citep{Ledig2017,Karras2019}.

\subsection{Self-supervised learning}

Closely related to representation learning is self-supervised learning that is today the state-of-the art methodology for obtaining informative and explanatory representations.
The appeal of self-supervised learning is that it does not require labeled data but uses for training a loss function that solely depends on the data itself.
In computer vision, for example, a common self-supervised learning task is to in-paint (or predict) a region that was cropped out from a given image~\citep{Pathak2016}. 
Since training is typically informed by the data and not a specific application, self-supervised learning fits naturally with representation learning where one seeks domain- or data-specific but task-independent representations.
The ability to use very large amounts of training data, which is usually much easier than in supervised training since no labels are required, also helps in most instances to significantly improve representations~\citep{Devlin2019,Radford2018,Zhai2021}.

Prominent examples of pretext tasks for image understanding include solving jigsaw puzzles~\citep{Noroozi2016b}, learning image rotations~\citep{Gidaris2018}, predicting color channels from grayscale images and vice-versa~\citep{Zhang2017CVPR}, or inpainting cropped out regions of an image~\citep{Pathak2016}. 
A early approach that has been used for representation learning is the denoising autoencoder~\citep{Vincent2010}. 
The work of~\cite{Mishra2016} is directly related to ours in the sense that they train a network to predict the temporal order of a video sequence using a triplet loss. In contrast, our approach relies on predicting the exact (categorical) temporal distance between two patches, not order, which we believe forces the network to learn more informative representations.

Recently, consistency-based methods have received considerable attention in the literature on self-supervised learning, e.g. in the form of contrastive loss functions or student-teacher methods. 
Since our work employs a pretext task, we will not discuss these methods but refer to~\cite{Le-Khac2020} for an overview.

\subsection{Machine Learning for the Geoscience}

Deep neural networks have become an important tool in the geosciences in the last years and will likely be relevant for a wide range of problems in the area in the future, e.g.~\cite{Toms2020,Schultz2021,Dueben2018,Balaji2022}.
\cite{Reichstein2019} pointed out the importance of spatio-temporal approaches to machine learning in the field. 
This implies the need for network architectures adapted to spatio-temporal data as well as suitable learning protocols and loss functions.
AtmoDist addresses the last aspect.

An early example of spatial representation learning in the geosciences is Tile2Vec~\citep{jean2019tile2vec} for remote sensing.
The work demonstrates that local tiles, or patches, can serve as analogues to words in geospatial data and that this allows for the adoption of ideas from natural language processing to geoscience.
Related is Space2Vec that can be considered as a multi-scale representation learning approach of geolocations.
A spatio-temporal machine learning approach is used in~\cite{Barnes2018} where a neural network is trained to predict the global year of a temperature field.
This is similar to AtmoDist where the training task is also the prediction of temporal information given a spatial field.
\cite{Barnes2018}, however, use their trained network for questions related to global warming whereas we are interested in representation learning.
To our knowledge, spatio-temporal representation learning, specifically for neural networks, as a means to improve downstream tasks has not been considered in atmospheric dynamics before, in contrast to, for instance, natural language processing~\citep{Devlin2019, Vaswani2017} or computer vision~\citep{Caron2021, dosovitskiy2020image}. 
In the past, dimensionality reduction techniques such as PCA, kernel-PCA, or MVU, have been used extensively to analyze atmospheric dynamics, e.g.~\citep{lima2009statistical, Hannachi2019, Mercer2012}. These can be seen as simple forms of representation learning since they also provide a data transformation to a coordinate system adapted to the input.
The need for more expressive representations than those obtained by these methods has been one of the main motivations behind deep neural networks, cf.~\citep{Bengio2013}.

A number of GANs dedicated to geoscience applications have been proposed in the literature, e.g.~\cite{klemmer2021sxl,zhu2020spatial,Klemmer2022,Stengel2020}.
Noteworthy is SPATE-GAN~\citep{Klemmer2022} that uses a custom metric for spatio-temporal autocorrelation and based on it determines an embedding loss. 
The authors demonstrate that this improves GAN performance across different data sets without changes to the neural network architecture of the generative model. 
It hence provides an alternative to AtmoDist proposed in our work.
We compare with~\cite{Stengel2020} which was focusing on impactful applications.

\section{Method}
\label{sec:architecture}

We perform self-supervised representation learning for atmospheric dynamics and derive a data-driven distance function for atmospheric states from it.
For this, we employ a siamese neural network~\citep{Chicco2021} and combine it with a novel, domain-specific spatio-temporal pretext task that derives from the theory of geophysical fluid dynamics.
Specifically, for a given temporal sequence of unlabelled atmospheric states, a neural network is trained to predict the temporal separation between two nearby ones.
For the predictions to be accurate, the network has to learn an internal representation that captured intrinsic properties of atmospheric flows, and hence provides feature spaces adapted to atmospheric dynamics.
For training we employ ERA5 reanalysis~\citep{era5}, which we consider a good approximation to observations.
An overview of the AtmoDist methodology is provided in Fig.~\ref{fig:atmodist}.

\subsection{Dataset and Preprocessing}
\label{sec:dataset}

We employ relative vorticity and divergence to represent an  atmospheric state.
The two scalar fields are equivalent to the wind velocity vector field, which is the most important dynamic variable and hence a good proxy for the overall state.
Our data is from model level 120 of ERA5, which corresponds approximately to pressure level $883 \textrm{hPa} \pm 85$, and a temporal resolution of three hours is used. 
Vorticity and divergence fields are obtained from the native spectral coefficients of ERA5 by mapping them onto a Gaussian grid with resolution $1280 \times 2560$ (we use~ pyshtools for this~\citep{pyshtools}).
The grids are subsequently sampled into patches of size $160 \times 160$, which corresponds approximately to $2500 \, \text{km} \times 2500 \, \text{km}$, with randomly selected centers. 
Following~\cite{Stengel2020}, we restrict the centers to $\pm 60^\circ$ latitude to avoid the severe distortions close to the poles.

\begin{figure}[t]
    \centering
    \includegraphics[width=0.8\textwidth]{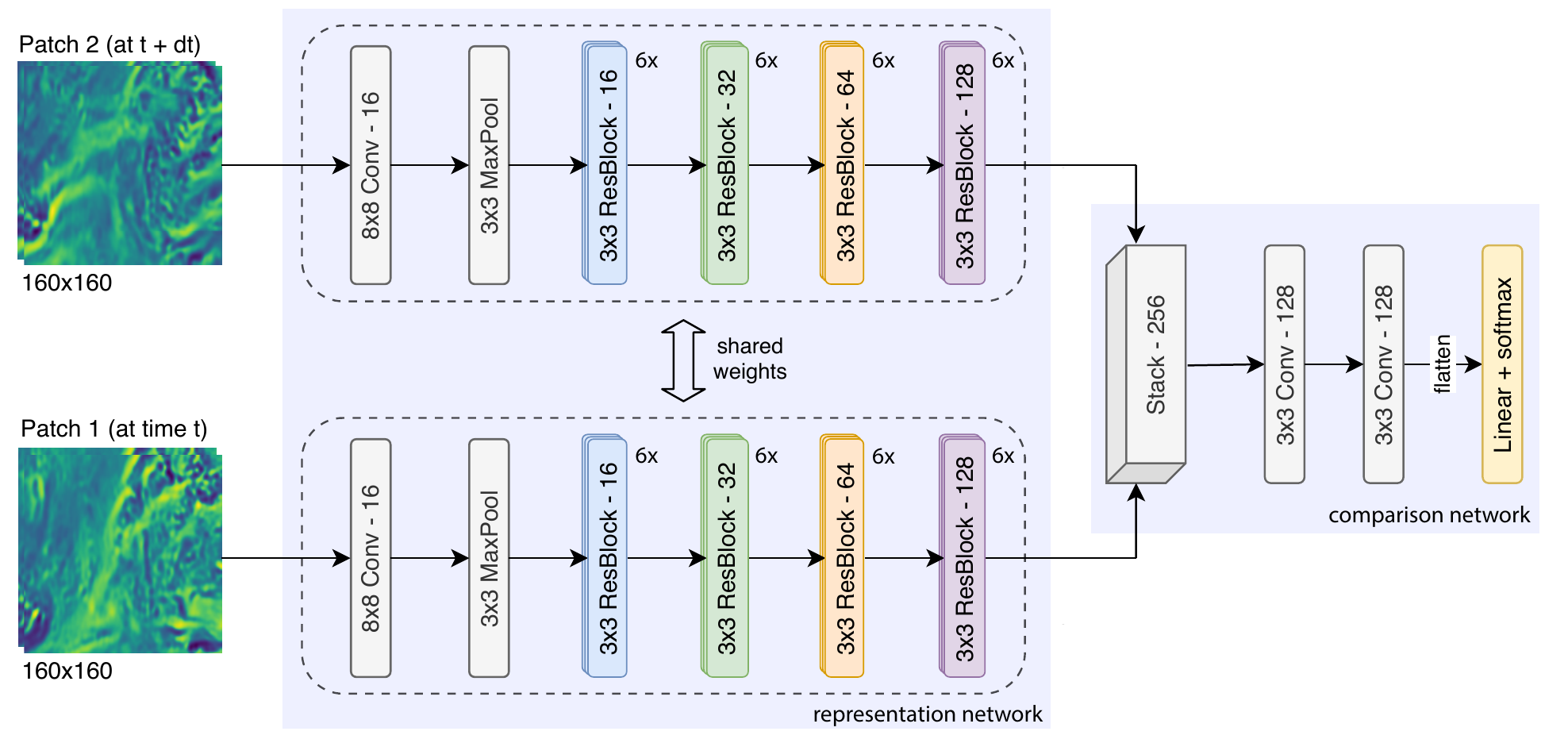}
    \caption{The AtmoDist network used for learning the pretext task. Numbers after layer names indicate the number of filters / feature maps of an operation. The comparison network is only required during training and can be discarded afterwards}
    \label{fig:network}
\end{figure}

\begin{table}[h]
    \centering
    \begin{tabular}{@{}lllll@{}}
    \toprule
    Task & AtmoDist & Super-resolution\\
    \midrule
    Dataset & ERA5 & ERA5\\
    Variables & Divergence, Vorticity & Divergence, Vorticity\\
    Model level & 120 & 120\\
    Training Period & 1979-1998 & 1979-1998\\
    Evaluation Period & 2000-2005 & 2000-2005\\
    Preprocessing & log-space & log-space\\
    Patch-size & $160 \times 160$ & $96 \times 96$\\
    Patches per timestep & 31 & 180\\
    Center between & $60\text{°N} - 60\text{°S}$ & $60\text{°N} - 60\text{°S}$\\
    Maximum latitude & $82.5\text{°N}$ / $82.5\text{°S}$ & $73.5\text{°N}$ / $73.5\text{°S}$\\
    Size (training) & $741 \text{GB}$ & $775 \text{GB}$ \\
    \bottomrule
    \end{tabular}
    \vspace{0.3cm}
    \caption{\centering Overview of the data used in this work}
    \label{tab:dataset}
\end{table}

We found that both vorticity and divergence roughly follow a zero-centered Laplace distribution.
This led to instabilities in particular in the training of the downstream task.
While clipping values larger than $70$ standard deviations was sufficient to stabilize training, this discards information about extreme events that is of particular relevance in many applications. 
We therefore apply an invertible log-transform to the input data in a preprocessing step and train and evaluate in the log-transformed space, see Appendix~\ref{app:preprocessing} for details.

Training is performed on $3$-hourly data from 1979 to 1998 (20 years) while the period from 2000 to 2005 is reserved for evaluation (6 years). 
This results in $58440 \, \times \, N_p$ spatial fields for the training and $17536 \, \times \, N_p$ fields for the evaluation set, where $N_p$ is the number of patches per global field of size $1280 \times 2560$. 
We used $N_p = 31$ in our experiments. The maximum time lag, i.e. the maximum temporal separation between spatial fields, was $\Delta t_{\textrm{max}} = 69 \, \text{h}$.
This is equivalent to $23$ categories for the training of the representation network.
An overview of the dataset is given in Table~\ref{tab:dataset}.

\subsection{Pretext Task}

Our pretext task is defined for a temporal sequence of spatial fields, e.g. atmospheric states from reanalysis or a simulation, and it defines a categorial loss function for self-supervised training.
The task is derived from the theory of geophysical fluid dynamics and motivated by the fact that the time evolution of an ideal barotropic fluid is described by a geodesic flow~\citep{Arnold1966,Ebin1970}. 
Since a geodesic flow is one of shortest distance, the temporal separation between two nearby states corresponds to an intrinsic distance between them.
As a spatio-temporal pretext task for learning a distance measure for atmospheric dynamics, we thus use the prediction of the temporal separation between close-by states.
More specifically, given two local patches of atmospheric states $X_{t_1}, X_{t_2}$ centered at the same spatial location but at different, nearby times $t_1$ and $t_2$, the task for the  neural network is to predict their temporal separation $\Delta t = t_2 - t_1 = n \cdot h_t$ given by a multiple of the time step $h_t$ ($3\mathrm{h}$ in our case).
The categorical label of a tuple $(X_{t_1},X_{t_2})$ of input patches, each consisting of the vorticity and divergence field at the respective time $t_k = k \cdot h_t$ for the patch region, is thus defined as the number of time steps $n$ in between them.
Following standard methodology for classification problems, for each training item $(X_{t_1},X_{t_2})$, our representation network predicts a probability distribution over the finite set of allowed values for $n$.
Training can thus be performed with cross-entropy loss, which is known to be highly effective. 

For a distance metric one expects $F(X_{t_1}, X_{t_2}) = F(X_{t_2}, X_{t_1})$.
However, we found that reversing the order of inputs results in prediction errors being reversed as well and training the network on randomly-ordered pairs did not prevent this behavior. As a consequence, we train the network using a fixed order, i.e. we only evaluate $F(X_{t_1}, X_{t_2})$ with $t_1 < t_2$.

\begin{figure}[t]
    \centering
    \includegraphics[width=0.38\textwidth]{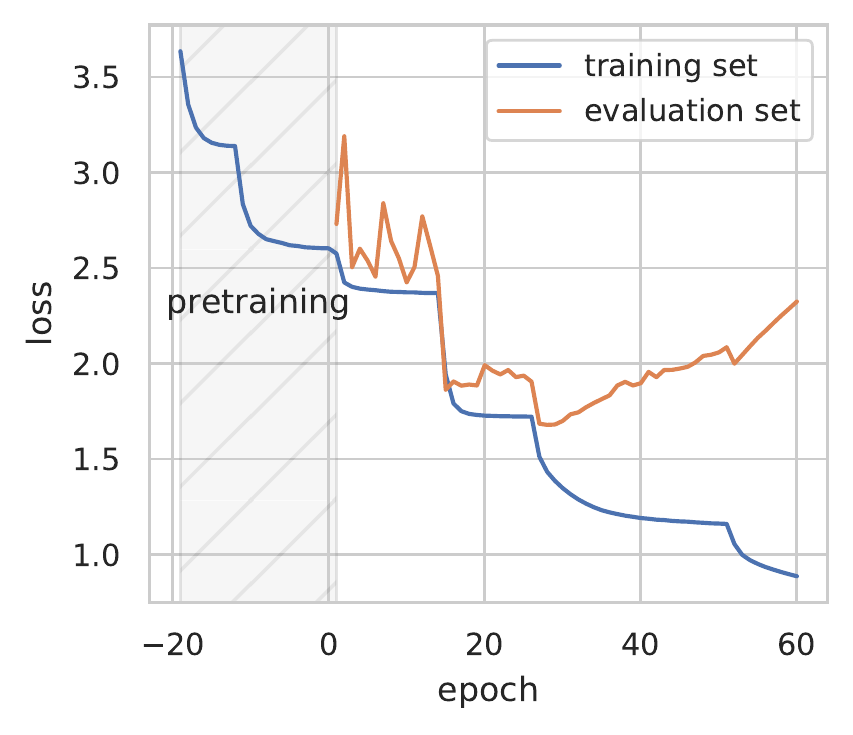}
    \qquad
    \includegraphics[width=0.38\textwidth]{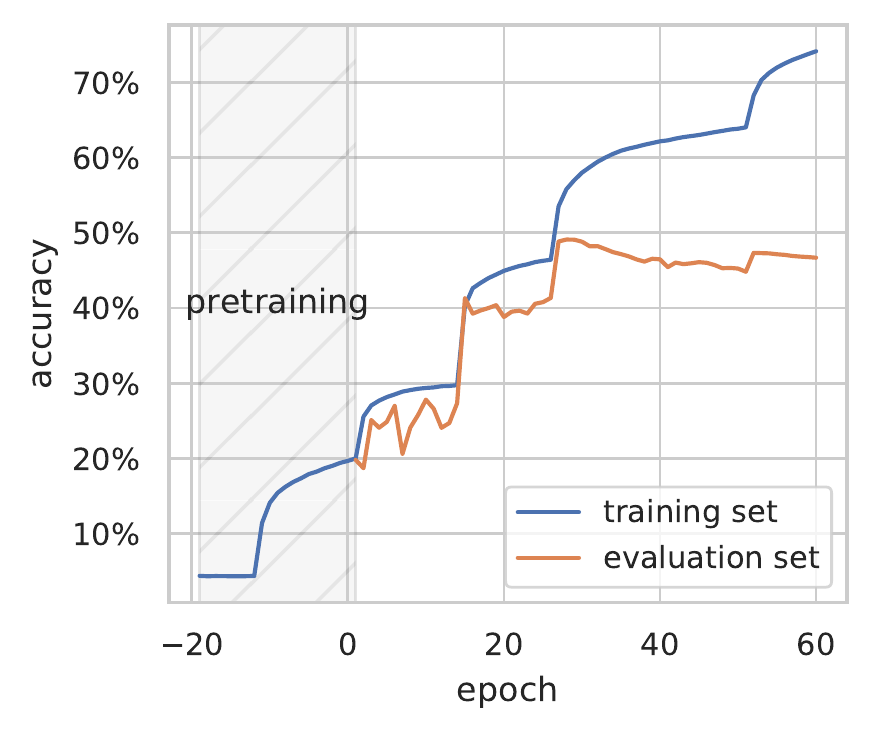}
    \caption{Loss (left) and Top-1 accuracy (right) during training calculated on both the training dataset (1979-1998) and the evaluation dataset (2000-2005). Drops in loss correspond to learning rate reductions. The best loss and accuracy are achieved in epoch 27 after which the network clearly begins to overfit}
    \label{fig:training_loss}
\end{figure}

\subsection{Neural Network Architecture}
\label{sec:network_architecture}

The architecture of the neural network we use for representation learning consists of two parts and is schematically depicted in Fig.~\ref{fig:network}.
The first part is the representation network.
It provides an encoder that maps an atmospheric field $X$ to its feature space representation $\mathcal{F}(X) \subseteq \R^N$.
Since both states $X_{t_k}$ of the tuple $(X_{t_1}, X_{t_2})$ that form a training item are used separately as input to the encoder, it is a siamese network~\citep{Chicco2021}.
The second part of our overall architecture is a tail or comparison network $T(\mathcal{F}(X_{t_1}), \mathcal{F}(X_{t_2}))$ that maps the tuple $(\mathcal{F}(X_{t_1}),\mathcal{F}(X_{t_2}))$ of representations to a probability density $p(\Delta t | X_{t_1}, X_{t_2})$ for their temporal separation $\Delta t = n \cdot h_t$. 
The representation and tail networks are trained simultaneously in an end-to-end manner.
After training, only the representation network is of relevance since its activations at the final layer provide the feature space $\mathcal{F}(X)$ for the input $X_{t}$ that defines the learned representation.
The use of activations at intermediate layers is also possible but was not considered in the present work.
Note that the tail network should be much smaller than the representation network to facilitate discriminative and explanatory representations.

The representation network follows a residual architecture~\citep{he2015deep} although with a slightly reduced number of feature maps compared to the standard configurations used in computer vision. 
It maps an input $X$ of size $2 \times 160 \times 160$ to a representation vector $\mathcal{F}(X)$ of size $128 \times 5 \times 5$.  
This corresponds to a compression rate of $16$.
The tail network is a simple convolutional network with a softmax layer at the end to obtain a discrete probability distribution. 
Both network together consist of $2,747,856$ parameters with $2,271,920$ in the encoder and $470,144$ in the tail network.

\subsection{Training}

We train AtmoDist on the dataset described in Sec.~\ref{sec:dataset} using stochastic gradient descent. 
Since training failed to converge in early experiments, we introduced a pre-training where we initially use only about $10\%$ of the data before switching to the full data set.
For further details of the training procedure, we refer to Appendix~\ref{app:training}.

As can be seen in Figure \ref{fig:training_loss}, with pre-training the training loss converges well although overfitting sets in from epoch 27 onwards.
Our experiments indicate that the overfitting results from using a fixed, pre-computed set of randomly sampled patches and it could likely be alleviated by sampling these dynamically during training. 
The noise seen in the evaluation loss is a consequence of the different training and evaluation behavior of the batch normalization layers.
While there exist methods to address this issue \citep{ioffe2017batch}, we found them insufficient in our case.
Instance normalization~\citep{ulyanov2017improved} or layer normalization~\citep{Ba2016} are viable alternatives that should be explored in the futures.


\subsection{Construction of AtmoDist metric}
\label{sec:method:atmodist_dist}


The final layer of the representation network provides an embedding $\mathcal{F}(X_t)$ of the vorticity and divergence fields, which together form $X_{t}$, into a feature space, cf. Fig.~\ref{fig:network}.
Although this representation can potentially be useful for many different applications, we employ it to define a domain-specific distance functions for atmospheric states.

The feature space representation $\mathcal{F}(X_t)$ is a tensor of size $128 \times 5 \times 5$ that we interpret as a vector, i.e. we consider $\mathcal{F}(X_t) \in \R^N$ with $N = 3200$.
We then define the AtmoDist metric $d(X_1, X_2)$ for two atmospheric states $X_1, X_2$ as
\begin{align}
    \label{eq:metric}
    d_{\mathrm{AtmoDist}}(X_1, X_2) = \frac{1}{N} \big\Vert \, \mathcal{F}(X_1) - \mathcal{F}(X_2) \big\Vert^2
\end{align}
where $\Vert \cdot \Vert$ denotes the standard $\ell_2$-norm. The $\ell_2$-norm is commonly used for the construction of metrics based on neural network activations~\citep{Gatys2016, Ledig2017}.
Other $\ell_p$-norms or weighted norms could potentially also be useful although preliminary experiments indicated that these provide results comparable to Eq.~\ref{eq:metric}.

\section{Evaluation}

\begin{figure}[t]
    \centering 
    \includegraphics[width=\textwidth]{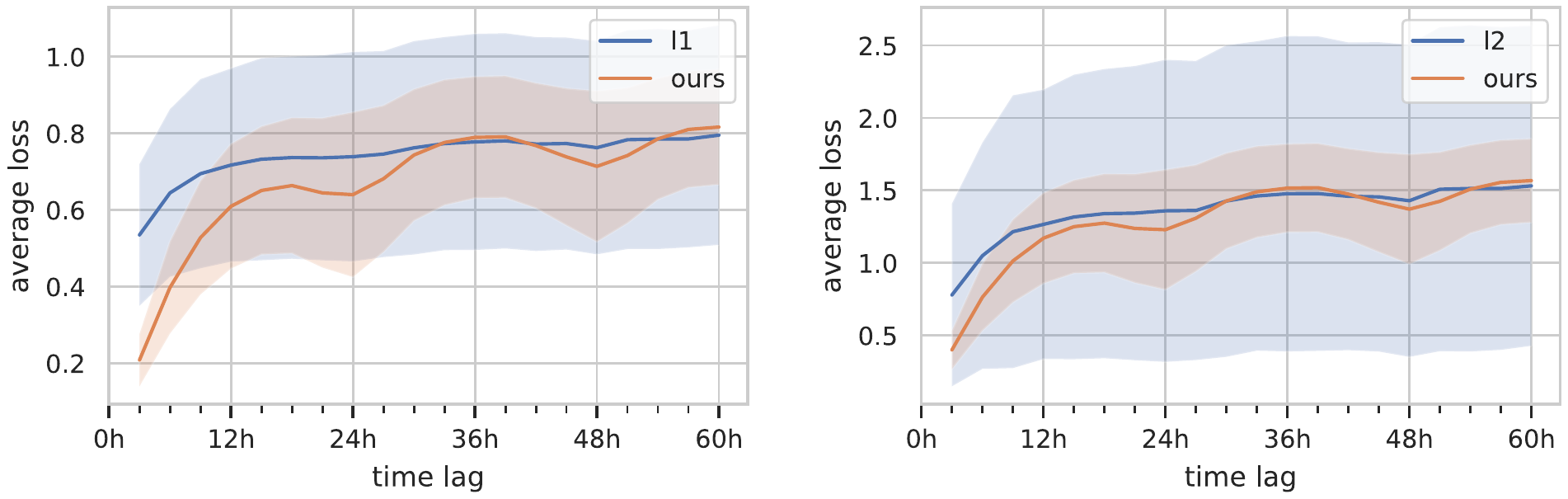}
     \caption{Mean $\ell_1$-norm (left) and mean $\ell_2$-norm (right) between samples that are a fixed time-interval apart, calculated on the training set. Shaded areas indicate standard deviation. For comparability, the AtmoDist distance has been normalized in each case with the method described in Appendix~\ref{app:scaling}. To give equal weight to divergence and vorticity, they have been normalized to zero mean and unit variance before calculating pixel-wise metrics.}
    \label{fig:metric_over_time}
\end{figure}

The evaluation of representation learning techniques usually employs a collection of downstream applications, since the embedding into the abstract and high-dimensional feature space is in itself rarely insightful.
To facilitate interpretation, one thereby typically relies on well known classification problems.
Simple techniques are also employed for the mapping from the representation to the prediction, e.g. a small neural network similar to our tail network, to indeed evaluate the representations and not any subsequent computations.

Unfortunately, standardized labeled benchmark datasets akin to MNIST~\citep{lecun1998gradient} or ImageNet~\citep{imagenet} currently do not exist  for atmospheric dynamics and it is their lack that inspired our self-supervised pretext task. 
We thus demonstrate the effectiveness of our representations using downscaling, i.e. super-resolution, and the interpolation of a partially missing field.
Both can be considered as standard problems and have been considered in a variety of previous works, e.g.~\citep{Groenke2020,Requena2019,Stengel2020,Meraner2020}.
For downscaling we build on the recent work by~\cite{Stengel2020} that provides a state-of-the-art GAN-based downscaling technique and, to facilitate a direct comparison, employ their implementation and replace only the $\ell_2$-norm in their code with the AtmoDist distance metric introduced in Sec.~\ref{sec:method:atmodist_dist}.
For missing data interpolation, we interpret it as a variant of inpainting and use a network inspired by those successful for the problem in computer vision.

As a baseline, we compare our learned representations against those of an autoencoder~\citep{Bengio2013}, one of the earliest representation learning methods. 
To facilitate a fair comparison, the encoder of the autoencoder is identical to representation network described in Sec.~\ref{sec:network_architecture}. The decoder is a mirrored version of the encoder, replacing downscaling convolutions with upscaling ones. For details, refer to Appendix~\ref{sec:appendix:autoencoder}. After training, the autoencoder is able to produce decent, yet overly-smooth, reconstructions as can be seen in Figure~\ref{fig:autoencoder_examples} in the Appendix.

Before we turn to the downstream applications, we begin with an intrinsic evaluation of the AtmoDist metric using the average distance between atmospheric states with a fixed temporal separation $\Delta t$.
Since this is close to the training task for AtmoDist, it provides a favorable setting for it.
Nonetheless, we believe that the comparison still provides useful insights on our work.

\subsection{Intrinsic evaluation of the AtmoDist distance}
 

\begin{figure}
    \centering
    \includegraphics[width=0.45\textwidth]{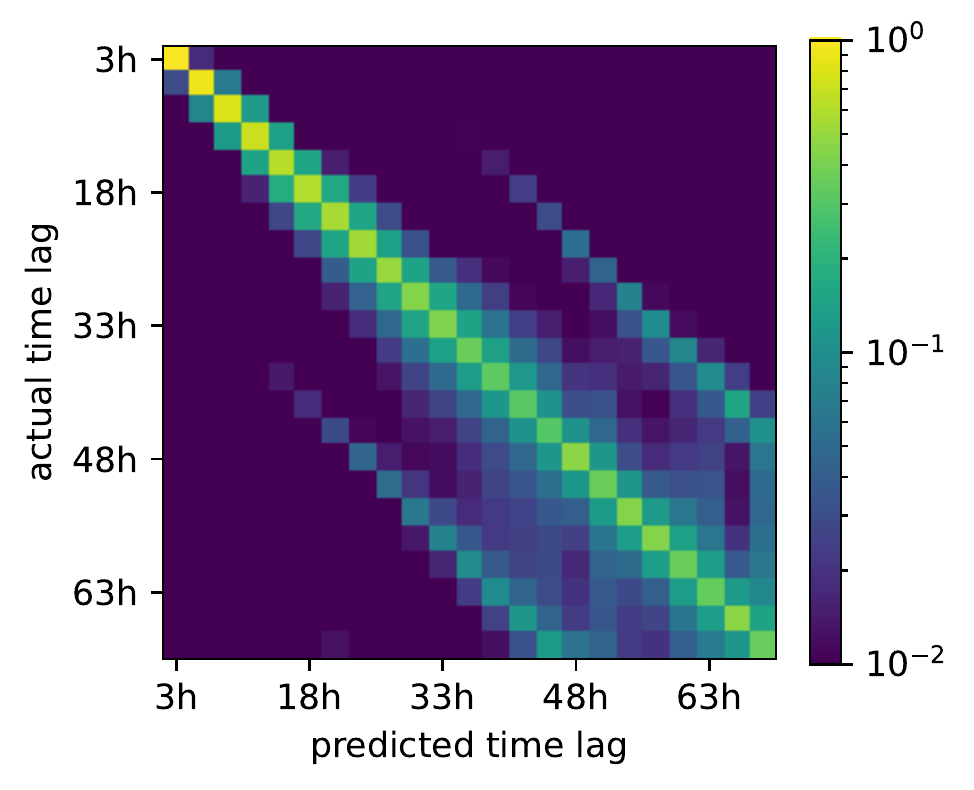}
    \caption{The confusion matrix shows the accuracy for the evaluation set as a function of predicted time lag and actual time lag. The side-diagonals indicate that AtmoDist is able to infer the exact time of the day for an atmospheric state with high precision solely based on a local patch of divergence and vorticity fields. We used a logarithmic scale to better highlight the side-diagonals}
    \label{fig:confusion}
\end{figure}

In order to obtain an intrinsic, application-independent evaluation of the AtmoDist distance in Eq.~\ref{eq:metric}, we determine it as a function of temporal separation $\Delta t$ between two atmospheric states $X_{t_1}$ and $X_{{t_2}}$. 
Note that although the training also employed $\Delta t$, the AtmoDist distance metric does no longer use the tail network and the computations are thus different than those during training.
Because of the quasi-chaotic nature of the atmosphere~\citep{Lorenz1969}, one expects that any distance measure for it will saturate when the decorrelation time has been reached.
To be effective, e.g., for machine learning applications, the distance between states should, however, dependent approximately linear on their temporal separation before the decorrelation time, at least in a statistical sense when a large number of pairs $X_{t_1}$ and $X_{{t_2}}$ for fixed $\Delta t$ is considered.

\paragraph{Comparison to $\ell_p$-norm,}
We compute $\ell_1$-norm, $\ell_2$-norm, and AtmoDist distance as a function of $\Delta t$ for all atmospheric states that form the training set for AtmoDist and report averaged distances for the different $\Delta t$.
As shown in Fig.~\ref{fig:metric_over_time}, the AtmoDist distance takes longer to saturate than mean $\ell_1$-norm and $\ell_2$-norm and increases more linearly. 
Also, its standard deviation is significantly smaller and AtmoDist hence provides more consistent distances.
Qualitatively similar results are obtained for SSIM~\citep{wang2004image} and PSNR, two popular metric in computer vision, and we report the results for these in Fig.~\ref{fig:ssim_psnr} in the appendix. 

\paragraph{Temporal behavior}
To obtain further insight into the temporal behavior of AtmoDist, we consider the confusion matrix as a functions of temporal separation $\Delta t$ when AtmoDist is used as in the training of the network, i.e. with the tail network.
Fig.~\ref{fig:confusion} confirms the expected behavior that predictions get less certain as $\Delta t$ increases and the states become less correlated. 
Interestingly, the emergence of sub-diagonals indicates that the network is able to infer the time of the day, i.e. the phase of Earth's rotation, with high precision, but it can for large $\Delta t$ no longer separate different days.

\begin{figure}
    \centering
    \includegraphics[width=\textwidth]{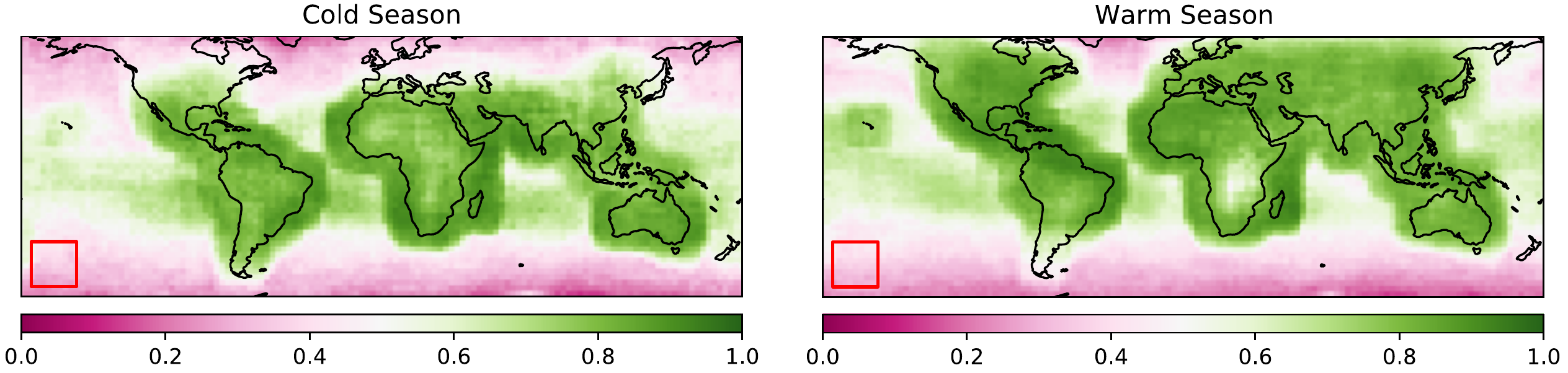}
    \caption{Accuracy of AtmoDist to correctly predict that two patches are $48 \, \text{h}$ apart as a function of space with an error margin of $3 \, \text{h}$ (i.e. $45 \, \text{h}$ and $51 \, \text{h}$ are also counted as correct prediction). The red rectangle in the lower left corner indicates the patch size used as input for the network.}
    \label{fig:spatial_errors}
\end{figure}

\paragraph{Spatial behavior} 
The predictability of atmospheric dynamics is not spatially and temporally homogeneous but has a strong dependence on the location as well as the season. 
One hence would expect that also the error of AtmoDist reflects these intrinsic atmospheric properties.
In Fig.~\ref{fig:spatial_errors} we show the spatial distribution of the error of AtmoDist, again in the setup used during training with the tail network.
As can been seen there, AtmoDist yields good predictions when evaluated near land but performance degrades drastically over the oceans.
Apparent in Fig.~\ref{fig:spatial_errors} is also a strong difference in predictability between the cold and warm season. 
This indicates that the model primarily focusses on detecting mesoscale convective activities and not on tracing Lagrangian coherent structures.

\subsection{Downscaling}
\label{sec:experiments:superres}

Downscaling, or super-resolution, is a classical problem in both climate science and computer vision. 
The objective is to obtain a high-resolution field $X^\text{hr}$ given only a low-resolution version $X^\text{lr}$ of it.
This problem is inherently ill-posed since a given $X^\text{lr}$ is compatible with a large number of valid high-resolution $X^\text{hr}$.
Despite this, state-of-the-art methods can often provide valid $X^\text{hr}$ whose statistics match those of the true fields. 
In the last years, in particular approaches based on generative adversarial networks (GAN)~\citep{Goodfellow2014} have become the de facto standard, e.g.~\cite{Stengel2020,Jiang2020,Klemmer2021b}.

\cite{Stengel2020} recently applied GAN-based super-resolution to wind and solar data in North America, demonstrating physically consistent results that outperform competing methods.
The authors build on the SRGAN from~\cite{Ledig2017} but instead of the VGG network~\citep{Simonyan2015} that was used as a representation-based content loss in the original work, \cite{Stengel2020} had to use an $\ell_2$-loss since no analogue for the atmosphere was available. 
Our work fills this gap and we demonstrate that the learned AtmoDist metric in Eq.~\ref{eq:metric} leads to significantly improved results for atmospheric downscaling.
The only modifications to the implementation from~\cite{Stengel2020} are a restriction to 4X super-resolution in our work (mainly due to the high computational costs for GAN training), incorporation of an improved initialization scheme for upscaling sub-pixel convolutions~\citep{aitken2017checkerboard}, as well as replacing transposed convolutions in the generator with regular ones as in the original SRGAN. 
We also do not use batch normalization in the generator, as suggested by~\cite{Stengel2020}. For both the $\ell_2$-based downscaling as well as the AtmoDist-based downscaling, the model is trained for $18$ epochs.
 

\begin{figure}
    \centering 
    \includegraphics[width=1.\textwidth]{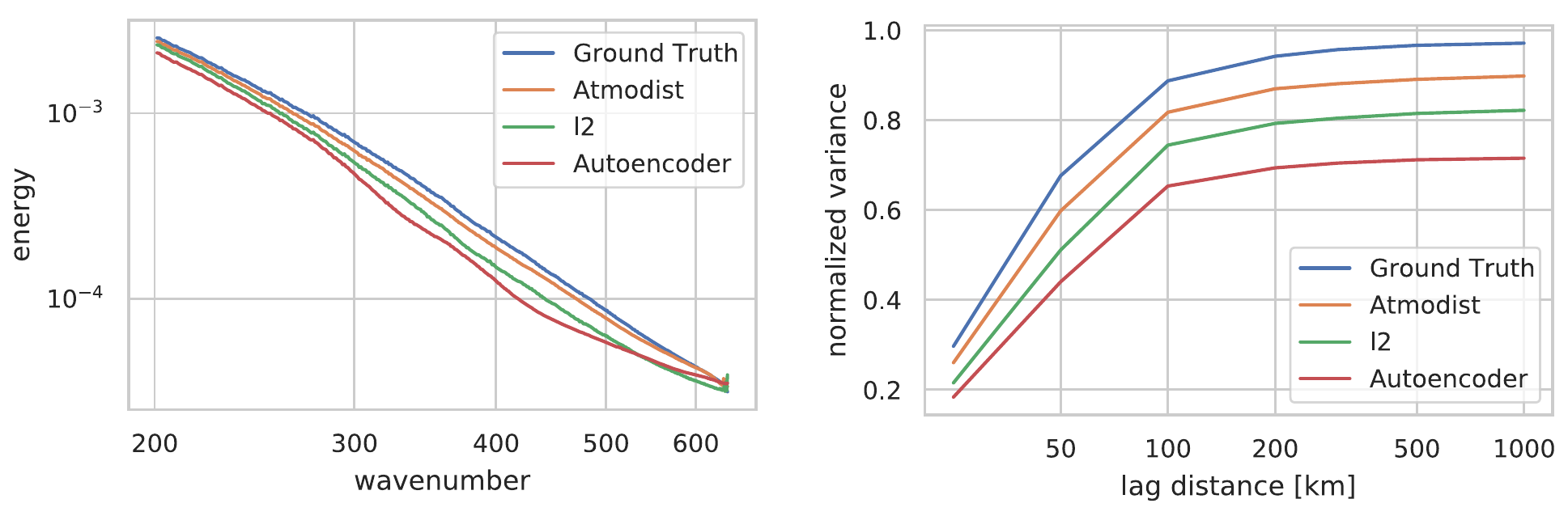}
    \caption{\emph{Left:} The energy spectrum from wavenumber 200 upwards averaged over the whole evaluation period. The spectra below wavenumber 200 are almost identical. The spectrum has been calculated by first converting divergence and vorticity to eastwardly and northwardly wind fields, and then evaluating the kinetic energy. \emph{Right:} Semivariogram of divergence}
    \label{fig:energy_spectrum}
\end{figure}

Downscaled images are shown in Fig.~\ref{fig:downscaling_examples1} and Fig.~\ref{fig:downscaling_examples2} in the appendix.
Qualitatively, the fields obtained with the AtmoDist metric look sharper than those with an $\ell_2$-loss. 
This overly smooth appearance with $\ell_2$-loss is a well known problem and one of the original motivations for learned content loss functions~\citep{Ledig2017}. 
In Fig.~\ref{fig:energy_spectrum} (left) we show the average energy spectrum of the downscaled fields. 
Also with respect to this measure, the AtmoDist metric provides significantly improved results and yields a spectrum very close to the ERA5 ground truth.
Following~\cite{Stengel2020}, we also compare the semivariogram of the downscaled fields
that measures the spatial variance of a field $f(x)$ as a function  of the lag distance $r$ ~\citep{matheron1963principles} (see Appendix~\ref{sec:appendix:semivariogram} for details on the calculation of the semivariogram).
As can be seen in Fig.~\ref{fig:energy_spectrum} (right) we find that our approach again captures the real geostatistics much better than an $\ell_2$-based downscaling.
The fields obtained from the autoencoder-based loss are visually of lower quality than with the other two loss functions and the same holds true for the quantitative evaluation metrics.

\begin{figure}
\centering
  \includegraphics[width=\textwidth]{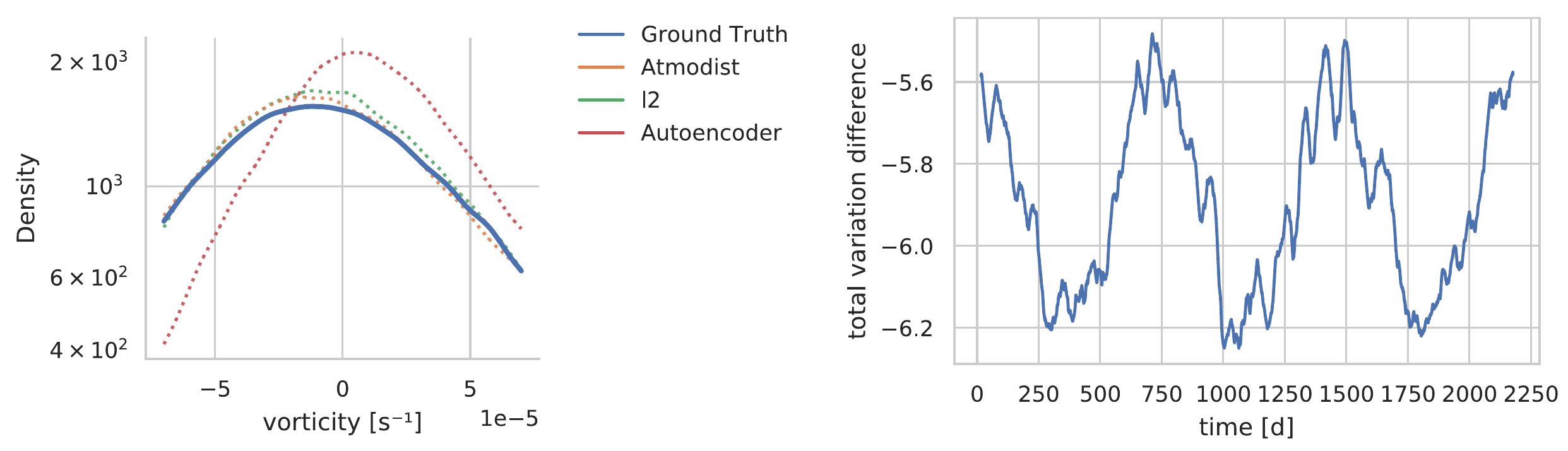}
\caption{\emph{Left: } Kernel density estimate of vorticity distribution at Milan (Italy). The $\ell_2$-based GAN achieves a Wasserstein distance of $5.3 \cdot 10^{-6}$ while our approach achieves a Wasserstein distance of $2.0 \cdot 10^{-6}$. The autoencoder-based GAN yields significant worse statistics. \emph{Right:} Reconstruction error measured as difference of total variation of divergence for the $\ell_2$-based super-resolution as a function of time. To highlight the oscillations, the errors have been smoothed by a $30\text{d}$ moving average. These oscillations are also present in the AtmoDist-based super-resolution, when comparing vorticity, or when the reconstruction error is measured using the $\ell_2$ norm}
  \label{fig:biennial_oscillation}
    \label{fig:milan_kde}
\end{figure}

Finally, we investigate local statistics for the GAN-based downscaling. 
In Fig.~\ref{fig:milan_kde} (left) we show these for vorticity.
The AtmoDist metric again improves the obtained results although a discrepancy to the ERA5 ground truth is still apparent. 
In Table~\ref{tab:local_stats}
\begin{table}[h!]
  \centering
    \begin{tabular}{@{}lllll@{}}
        \toprule
        Variable & Better & Equal & Worse \\
        \midrule
        Divergence & 102 & 12 & 36 \\
        Vorticity & 90 & 11 & 49 \\
        \bottomrule
    \end{tabular}  
    \vspace{0.25cm}  
    \caption{\centering Better/worse scores for local statistics of GAN-based super-resolution.}
    \label{tab:local_stats}
\end{table}
we report better/worse scores for AtmoDist-based downscaling and those using the $\ell_2$-loss for the Wasserstein-1 distance  calculated on the empirical distributions (akin to those in Fig.~\ref{fig:milan_kde}) for 150 randomly-selected, quasi-uniformly distributed cities.
    A location is thereby scored as better if the Wasserstein-1 distance of the $\ell_2$-based super-resolution exceeds $10\%$ of the Wasserstein-1 distance of our approach, and as worse in the opposite case. If neither is the case, i.e. both approaches have a comparable error, the location is scored as equal.
We find that for divergence we achieve better Wasserstein-1 distances in 102 out of 150 locations while only being worse in 36 out of 150.
Similar results are obtained for vorticity.


\begin{figure}
    \centering
    \includegraphics[width=\textwidth]{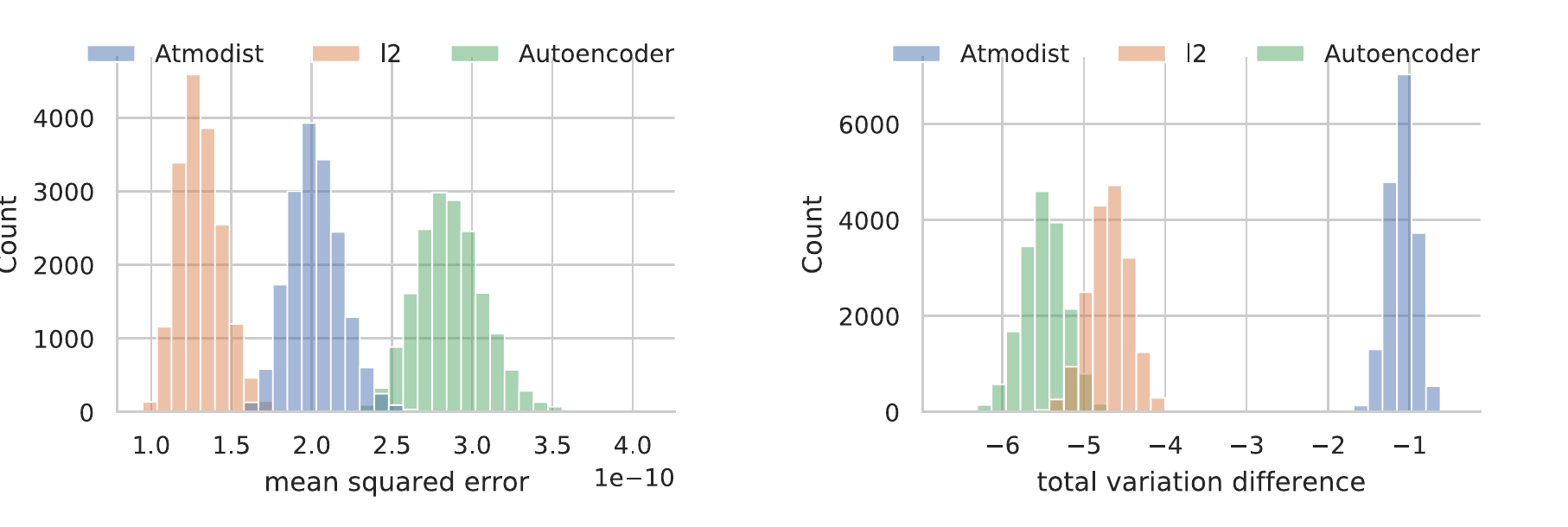}
    \caption{Histogram of reconstruction errors measured in $\ell_2$ norm~(left) and difference of total~variation~(right) for relative vorticity. We define the difference of total variation between the original field $f$ and its super-resolved approximation $g$ as $d_\text{tv}(f, g) = \int_\mathcal{D} \left|\nabla f(x)\right| - \left|\nabla g(x)\right| dx$. Values closer to zero are better. Despite performing better with regards to the $\ell_2$ reconstruction error, the $\ell_2$-based super-resolution performs worse with regards to the difference of total variation. Notice that the approach by Stengel et al. specifically minimizes the $\ell_2$ reconstruction error. Interestingly, all three approaches have solely negative total variation differences, implying that the super-resolved fields are overly smooth compared to the ground~truth fields. Similar results are obtained for divergence.}
    \label{fig:reconstruction_error}
\end{figure}

\paragraph{Biennial oscillations}
In Fig.~\ref{fig:biennial_oscillation} (right) we show the downscaling error for divergence over the six year evaluation period. 
Clearly visible is an oscillation in the error with a period of approximately two years, which exist also for vorticity and when $\ell_2$-loss is used.
It is likely that these oscillations are related to the quasi-biennial oscillation (QBO)~\citep{Baldwin2001} and thus reflect intrinsic changes in the predicability in the atmosphere.
We leave a further investigation of the effect of the QBO on AtmoDist to future work.

\subsection{Reconstruction of Partially Occluded Fields}

In the atmospheric sciences, the complete reconstruction of partially occluded or missing fields, for example because of clouds, is an important problem, e.g.~\citep{Meraner2020}.
It appears in a similar form in computer vision as inpainting.

To further evaluate the performance of AtmoDist, we also use it as loss for this problem and compare again against the $\ell_2$-loss.
For simplicity, we use again ERA5 divergence and vorticity fields as data set and artifically add occlusion by cropping out $40 \times 40$ region centrally from the $160 \times 160$ patches used in the training of the representation network.

As network for the inpaiting, we choose the identical architecture as for the autoencoder (see Appendix~\ref{sec:appendix:autoencoder}). 
The $160 \times 160$ image with cropped-out center is passed as input and the network outputs an equally-sized image. From that output, only the central region is considered when calculating the loss during training. 
Details are presented in Appendix~\ref{sec:appendix:inpainting}.

Figure~\ref{fig:inpainting_example} shows reconstructed fields for all three loss functions. The AtmoDist-based reconstruction produces the most detailed field although it suffers from some blocking artifacts. 
The $\ell_2$-based reconstruction is overly smooth and has homogenous structures instead of fine details. 
The autoencoder-based field is even smoother than the reconstruction based on the $\ell_2$ loss. 
In fact, we had to tune our training procedure by switching to Adam and lowering the learning rate to prevent this model from generating constant fields. 
This was not necessary for either the $\ell_2$-based or AtmoDist-based reconstructions.
Semiovariograms for the reconstructions are shown in Fig.~\ref{fig:inpaining_semivariograms} in the appendix. 
These also verify that AtmoDist provides reconstructions with more realistic fine scale details.
The average $\ell_2$ norms for AtmoDist, $\ell_2$-loss, and autoencoder loss are $0.62$, $0.50$, $0.88$. Again a clearly better performance of AtmoDist compared to the autoencoder can be observed.

\begin{figure}
  \includegraphics[width=\textwidth]{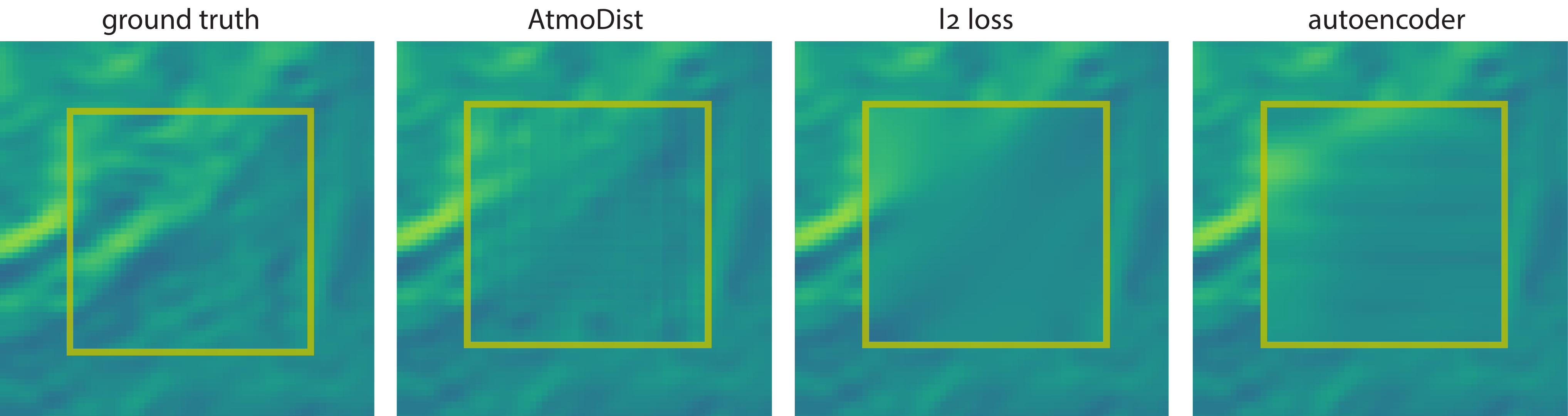}
  \caption{Interpolated vorticity fields (center, in the yellow square) for deleted regions for (form left to right) ground truth, AtmoDist, $\ell_2$-norm, and autoencoder. Both $\ell_2$-loss and autoencoder-loss produce overly smooth results. The AtmoDist-based reconstruction captures more of the higher-frequency features present in the data although it suffers from some blocking artifacts. The horizontal artifacts for the autoencoder occurred frequently in our results}
  \label{fig:inpainting_example}
\end{figure}

\subsection{Ablation study}

We performed an ablation study to better understand the effect of the maximum temporal separation $\Delta t_{\mathrm{max}}$ on the performance of AtmoDist.
If $\Delta t_{\mathrm{max}}$ is chosen too small, the pretext task might become too easy and a low training error might be achieved with sub-optimal representations. 
If $\Delta t_{\mathrm{max}}$ is chosen too large, the task might, however, become too difficult and also lead to representations that do not capture the desired effects.
We thus trained AtmoDist with $\Delta t_{\mathrm{max}} = \left\{45 \, \text{h}, 69 \, \text{h}, 93 \, \text{h} \right\}$ on a reduced dataset with only $66\%$ of the original size.
Afterwards, we train three SRGAN models, one for each maximum temporal separation, for $9$ epochs using the same hyper-parameters and dataset as in the original downscaling experiment. 

\begin{figure} 
    \centering
    \includegraphics[width=\textwidth]{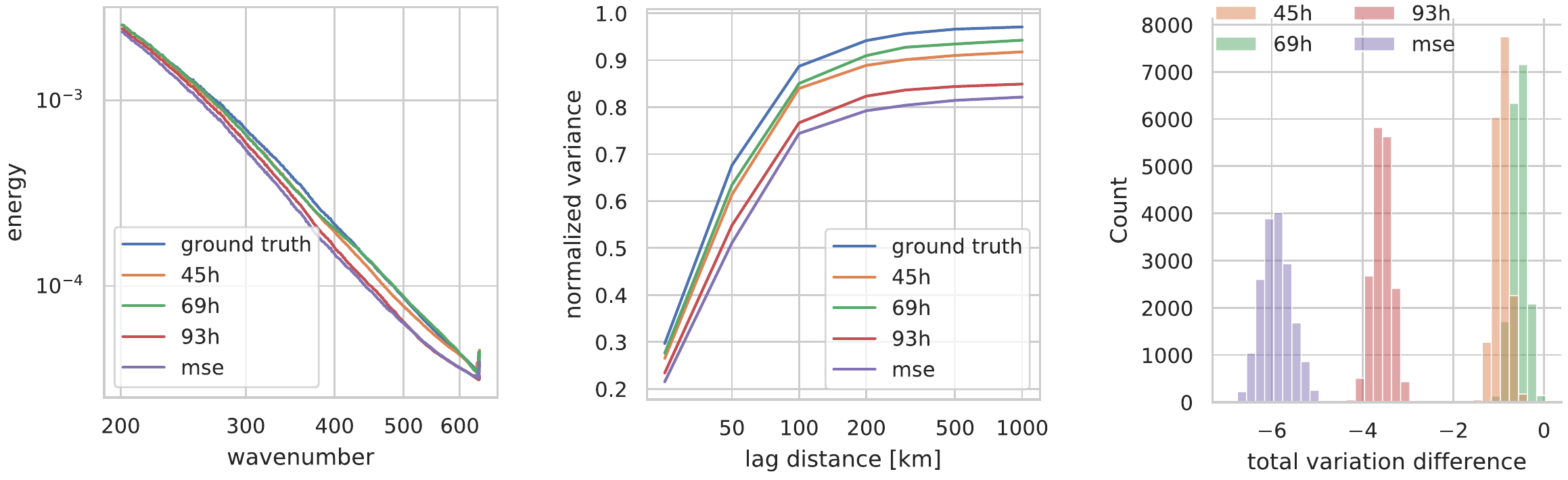}
    \caption{The energy spectrum (left), semivariogram (center), and distribution of total variation difference errors (right) for models trained with different maximum $\Delta t_{\mathrm{max}}$ for our ablation study. The semivariogram and error distributions are calculated on divergence, but qualitative similar results are obtained for vorticity.}
    \label{fig:ablation}
\end{figure}

Results for the energy spectrum, semivariogram, and reconstruction errors are shown in Figure~\ref{fig:ablation}. 
We find that with $\Delta t_{\mathrm{max}} = 69 \, \text{h}$ the downscaling performs slightly better than with $\Delta t_{\mathrm{max}} = 45 \, \text{h}$ with respect to all three metrics.
For $\Delta t_{\mathrm{max}} = 93 \, \text{h}$, the model performs significantly worse than the other two, implying that past a certain threshold performance begins to degrade. 
Notably, all three models outperform the $\ell_2$-based downscaling model even though the representations networks have been trained with less data as in the main experiment.

\section{Conclusion and Future Work}
\label{sec:conclusion}

We have presented AtmoDist, a representation learning approach for atmospheric dynamics.
It is based on a novel spatio-temporal pretext task designed for atmospheric dynamics that is applicable to a wide range of different fields. 
We used the representations learned by AtmoDist to introduce a data-driven metric for atmospheric states and showed that it improves the state-of-the-art for downscaling when used as loss function there.
For the reconstruction of missing field data it also led to more detailed and more realistic results.
Surprisingly, AtmoDist improved the performance even for local statistics, although locality played no role in the pretext task.  
These results validate the quality of our learned representations.

\paragraph{Possible extensions of AtmoDist}
We believe that different extensions of AtmoDist should be explored in the future. 
One possible direction is the use of a contrastive loss instead of our current pretext task.
For this, samples within a certain temporal distance to each other can be used as positive pairs and samples above that threshold as negative ones, akin to word2vec~\citep{Mikolov2013}.
However, we believe that predicting the exact time lag between two atmospheric states provides a much more challenging task and hence a better training signal than solely predicting if two states are within a certain distance to each other. Exploring a triplet loss~\citep{hoffer2015deep} is another interesting direction. 

We also want to explore other downstream tasks, e.g. the classification and prediction of hurricanes~\citep{Prabhat2021} or extreme events~\citep{Racah2017}.
Interesting would also be to explore transfer learning for AtmoDist, e.g. to train on historical data and then adapt to a regime with significant $\mathrm{CO}_2$ forcing, similar to~\cite{Barnes2018}.
This could be explored with simulation data, which can be used to train AtmoDist without modifications.

We employed only divergence and vorticity and a single vertical layer in AtmoDist. In the future, we want to validate our approach using additional variables, e.g. those appearing in the primitive equations, and with more vertical layers. 
It is also likely that better representations can be obtained when not only a single time step but a temporal window of nearby states is provided to the network. 

\paragraph{Outlook}
We consider AtmoDist as a first proof-of-concept for the utility of representation learning for analyzing, understanding and improving applications in the context of weather and climate dynamics.


Representation learning in computer vision relies heavily on data augmentation, e.g.~\cite{Chen2020MLR,Caron2021}.
While this is a well understood subject for natural images, the same does not hold true for atmospheric and more general climate dynamics data. 
Compared to computer vision, many more physical constraints have to be considered.
We hence believe that the design and validation of novel data augmentations is an important direction for future work.

Another currently unexplored research direction is representation learning using (unlabeled) simulation data.
For example, one could perform pretraining on the very large amounts of simulation data that are available from CIMP runs~\citep{Eyring2016} and use fine-tuning~\citep{Devlin2019}, transfer learning, or domain adaptation to derive a network that is well suited for observational data.
Another interesting direction is to compare representations obtained for reanalysis and simulation data, which has the potential to provide insights into subtle biases that persist in simulations.

Our current work focused on improving downstream applications using representation learning.
However, we believe that it also has the potential to provide new insights into the physical processes in the atmosphere, analogous to how tools such as proper orthogonal decompositions helped to analyze the physics in the past.
In our opinion, in particular attention-based network architectures, such as transfomers~\citep{Vaswani2017}, provide a promising approach for this.

\paragraph{Acknowledgments}
We gratefully acknowledge discussions with the participants of the workshop \emph{Machine Learning and the Physics of Climate} at the Kavli Institute of Theoretical Physics in Santa Barbara that helped to shaped our overall understanding of the potential of representation learning for weather and climate dynamics.

\paragraph{Funding Statement}
Funded by the Deutsche Forschungsgemeinschaft (DFG, German Research Foundation) – Project-ID 422037413 – TRR 287.

\paragraph{Competing Interests}
None. 

\paragraph{Data Availability Statement}
Our code is made available at \url{https://github.com/sehoffmann/AtmoDist}.

\paragraph{Ethical Standards}
The research meets all ethical guidelines, including adherence to the legal requirements of the study country.

\paragraph{Author Contributions}
Conceptualization: S.H.; C.L. Methodology: S.H; C.L. Data curation: S.H. Data visualisation: S.H. Writing original draft: S.H.; C.L. All authors approved the final submitted draft.

\paragraph{Supplementary Material}
No supplementary material. Code is available online.

\bibliography{paper}

\begin{appendix}
\section{Appendix.}\label{appendixA}

\subsection{Preprocessing}
\label{app:preprocessing}
 
Divergence and vorticity are transformed in a preprocessing step by $y = f(g(h(x)))$ where
\begin{align}
    \label{eq:pre}
    y = f(w) = \frac{w - \mu_2}{\sigma_2} \qquad
    w = g(z) = \sign(z) \log(1 + \alpha \left|z\right|) \qquad
    z = h(x) = \frac{x - \mu_1}{\sigma_1} \qquad
\end{align}
and which is applied element-wise and independently for vorticity and divergence.
Here $\mu_1$ and $\sigma_1$ denote the mean and standard deviation of the corresponding fields, respectively, while $\mu_2$ and $\sigma_2$ denote the mean and standard deviation of the log-transformed field $w$. All moments are calculated across the training dataset and are shown in Table~\ref{tab:means}. 
The parameter $\alpha$ controls the strength by which the dynamic range at the tails of the distribution is compressed. We found that $\alpha = 0.2$ is sufficient to stabilize training while it avoids an aggressive compression of the original data. 
Notice that the log function behaves approximately linear around 1, thus leaving small values almost unaffected.

\begin{table}[h!]
\centering
\begin{tabular}{@{}lllll@{}}
\toprule
Variable & $\mu_1$ & $\sigma_1$ & $\mu_2$ & $\sigma_2$ \\
\midrule
Divergence       & $1.9464334 \times 10^{-8}$ & $2.8568757 \times 10^{-5}$ & $8.821452 \times 10^{-4}$ & $1.5794525 \times 10^{-1}$\\
(Rel.) Vorticity & $2.0547947 \times 10^{-7}$ & $5.0819430 \times 10^{-5}$  & $3.2483143 \times 10^{-4}$ & $1.6044095 \times 10^{-1}$\\
\bottomrule
\end{tabular}
\vspace{0.2cm}
\caption{Mean and standard deviations calculated on the training dataset (1979-1998) on model level 120 for divergence and relative vorticity.}
\label{tab:means}
\end{table}

\subsection{AtmoDist Training}
\label{app:training}
The AtmoDist network is trained using standard stochastic gradient descent with momentum $\beta = 0.9$ and an initial learning rate of $\eta = 10^{-1}$. If training encounters a plateau, the learning rate is reduced by an order of magnitude to a minimum of $\eta_\text{min} = 10^{-5}$. Additionally, gradient clipping is employed, ensuring that the $l_2$-norm of the gradient does not exceed $G_\text{max} = 5.0$. Finally, to counteract overfitting, weight decay of $10^{-4}$ is used.

\begin{figure}[t]
  \centering
  \includegraphics[width=0.45\textwidth]{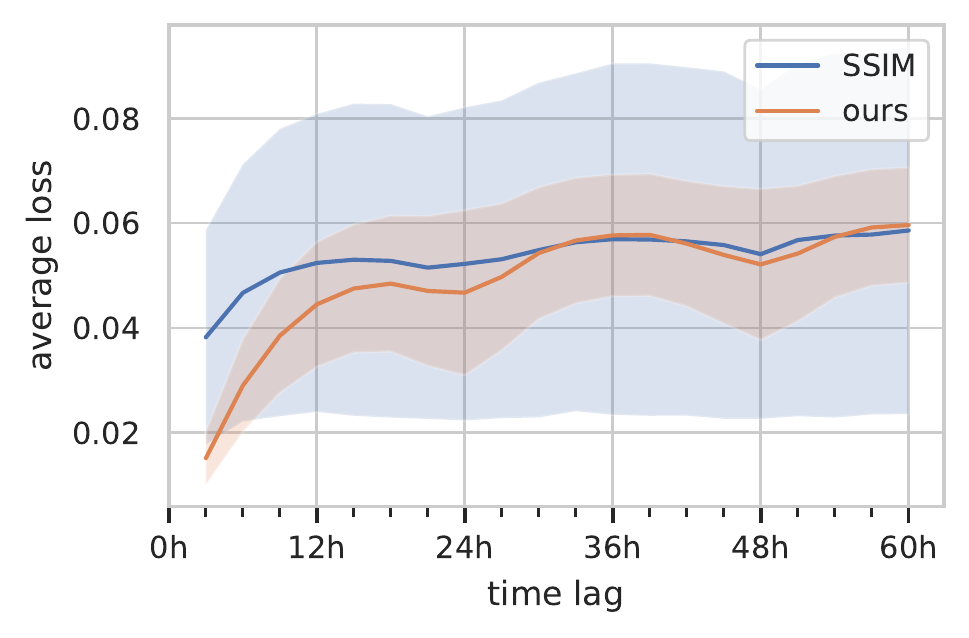}
  \qquad
  \includegraphics[width=0.45\textwidth]{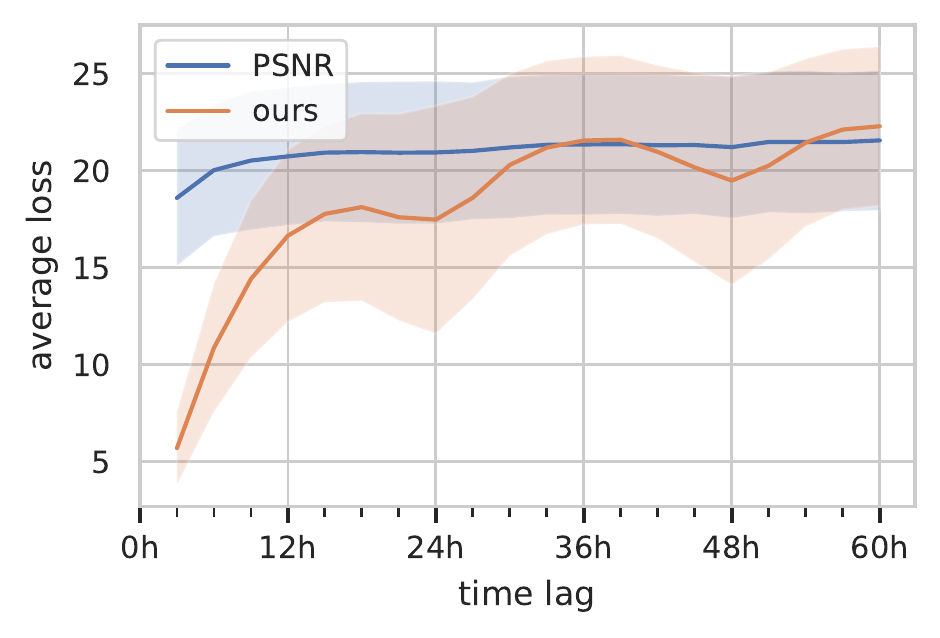}
  \caption{Mean SSIM and PSNR as a function of the temporal separation $\Delta t$. Since in both cases higher quantities indicate more similarity between samples, we apply the following transformations to make the plots comparable to Fig.~\ref{fig:metric_over_time}: SSIM: $y = 1 - \left(1 + \text{SSIM} (X_{t_1}, X_{t_2})\right) / 2 $; 
  PSNR: $y = 50 \, \text{dB} - \text{PSNR}(X_{t_1}, X_{t_2})$}
  \label{fig:ssim_psnr}
\end{figure}

Despite the network converging on lower resolutions in preliminary experiments, once we trained on 160x160 patches at native resolution (1280x2560) the network failed to converge.
We hypothesize that the issue is the difficulty of the pretext task combined with an initial lack of discerning features. We thus employ a pre-training scheme inspired by curriculum~learning~\citep{curriculum}. 
More specifically, we initially train the network only on about $10\%$ of the data so that it can first focus on solving the task there. 
After 20 epochs, we then reset the learning rate to $\eta=10^{-1}$ and start training on the whole dataset. 

\subsection{Scaling the loss function}
\label{app:scaling}

To ensure that downscaling with $\ell_2$-loss and the AtmoDist metric exhibit the same training dynamics, we normalize our loss function. This is particularly important with respect to the $\alpha_\text{adv}$ parameter which controls the trade-off between content-loss and adversarial-loss in SRGAN~\citep{Ledig2017}. 
The same procedure is also applied to the loss function derived from the autoencoder.

\begin{figure}
  \includegraphics[width=\textwidth]{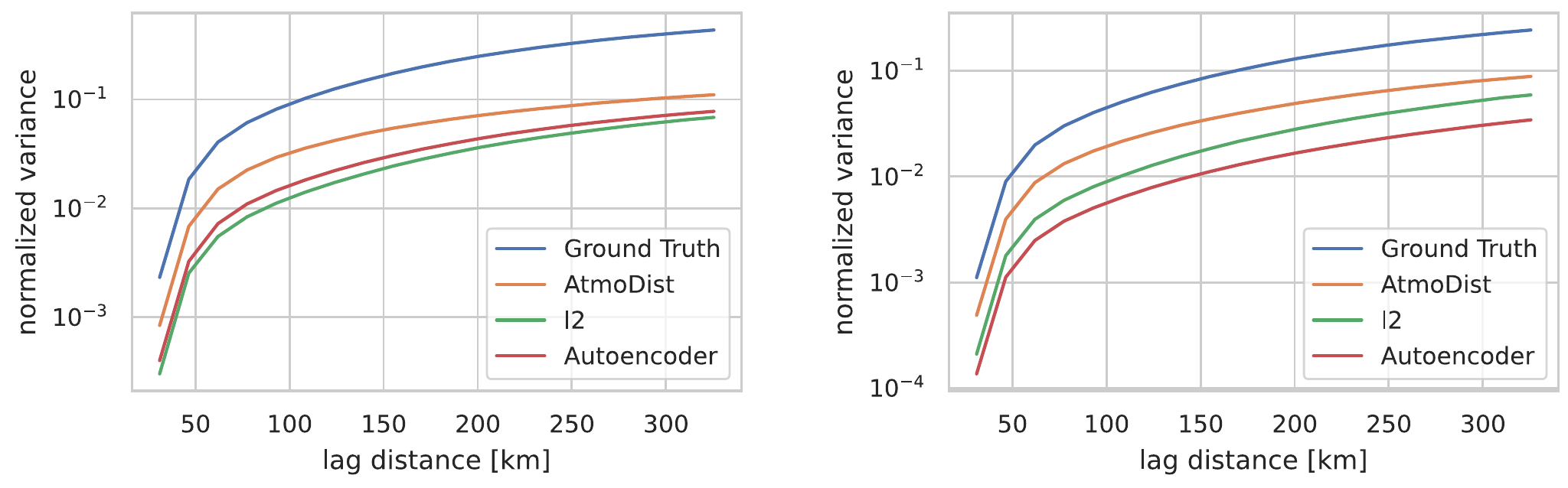}
  \caption{Semivariograms for the reconstruction of partially occluded fields for divergence (left) and vorticity (right). We hypothesis that the autoencoder performs better on divergence than the $\ell_2$-loss while the roles are interchanged for vorticity due to the larger high frequency content of divergence}
  \label{fig:inpaining_semivariograms}
\end{figure}

We hypothesize that due to the chaotic dynamics of the atmosphere, any loss function should on average converge to a specific level after a certain time period (ignoring daily and annual oscillations). Thus, we normalize our content-loss by ensuring that the equilibrium levels are roughly the same in terms of least squares by solving the following optimization problem for the scaling factor $\alpha_\text{cnt}$
\begin{equation}
\underset{\alpha_\text{cnt} \in \mathbb{R}}{\text{minimize}} \ \sum_{t={\lfloor}N/2{\rfloor}}^N (\alpha_\text{cnt} \mathbf{c}_t - \mathbf{m}_t)^2
\end{equation}
where $\mathbf{c}_t$ denote the average AtmoDist distance of samples that are $\Delta t$ apart and $\mathbf{m}_t$ their average $\ell_2$ distance. 
It is easy to verify that the above optimization problem has the unique solution
\begin{equation}
  \alpha_\text{cnt} = \frac{\sum_{t={\lfloor}N/2{\rfloor}}^N \mathbf{c}_t \mathbf{m}_t} {\sum_{t={\lfloor}N/2{\rfloor}}^N \mathbf{c}_t^2} .
\end{equation}

\subsection{Autoencoder Architecture}
\label{sec:appendix:autoencoder}

The autoencoder takes as input a divergence and vorticity field of size $160 \times 160$. It consists of an encoder part that compresses the input field to a suitable representation, and a decoder that takes the representation and reconstructs the original field from it.

The encoder is identical to the representation network used for the AtmoDist task. The decoder is a mirrored version of the encoder where downscaling convolutions were replaced by upscaling ones. Upscaling is done by bilinear interpolation followed by a standard residual block.

In the middle, i.e. in-between encoder and decoder, no feed-forward layer is used. It would have contained the majority of parameters of the overall network and thus potentially also of its capacity. Instead, we use an approach inspired by~\cite{mlpmixer} to ensure that information can propagate between each spatial position.

First, the $H \times \text W \times C$ feature map of the last encoder layer, where $H,W,C$ denote height, width, and number of channels respectively, is interpreted as $C$ vectors of dimensionality $H \cdot W$. Each vector is then transformed by a feed-forward layer, mixing information spatially for each channel. Afterwards, the same procedure is repeated for the $H \cdot W$ vectors of dimensionality $C$ to mix information channel-wise as well. This approach allows us to propagate information globally without bloating the size of the network in a significant way.

The autoencoder is trained in the same way as the AtmoDist representation network, cf. Appendix~\ref{app:training}, except that no pretraining is used.

\subsection{Inpainting Training}
\label{sec:appendix:inpainting}

\begin{figure}
  \centering
  \includegraphics[width=0.27\textwidth]{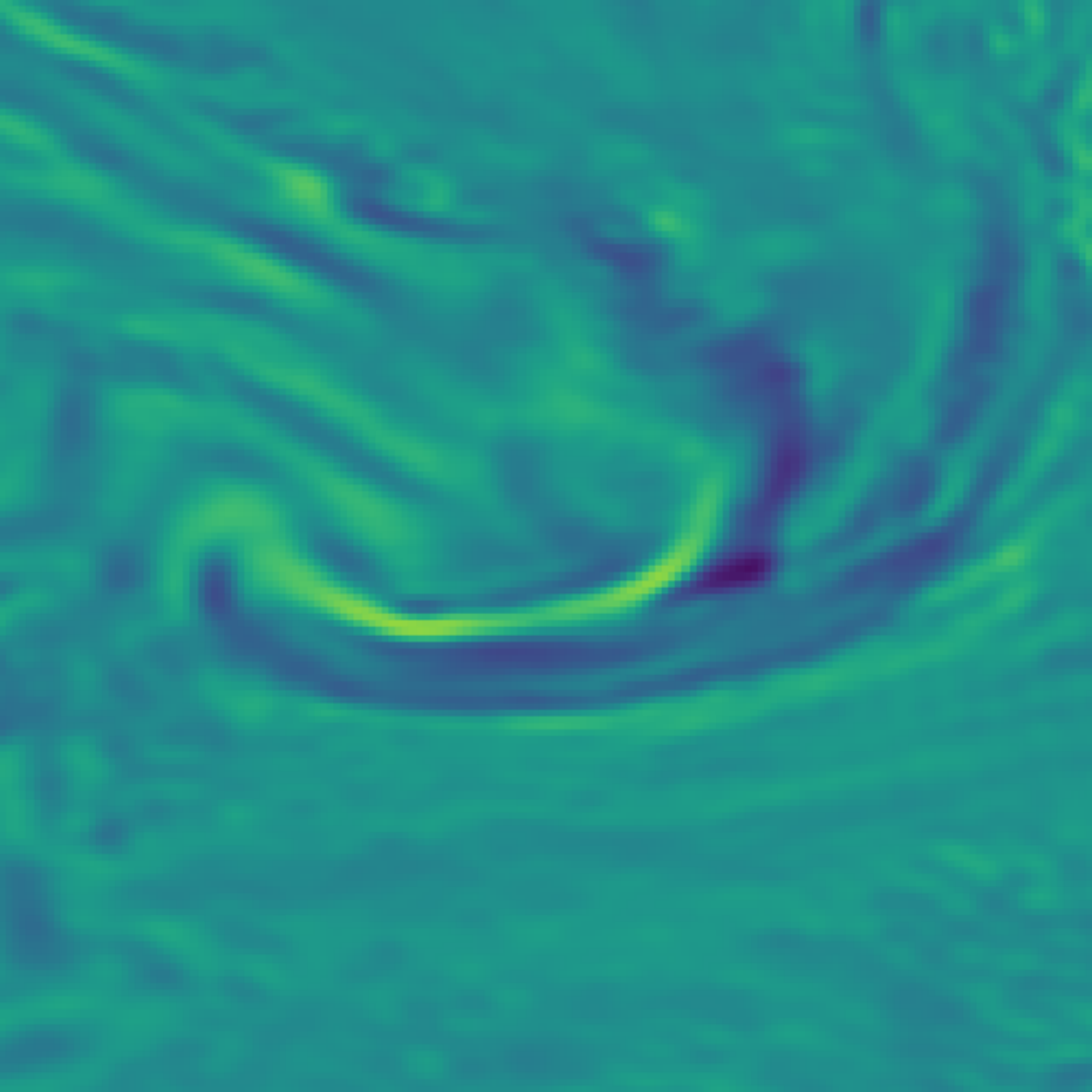}
  \quad
  \includegraphics[width=0.27\textwidth]{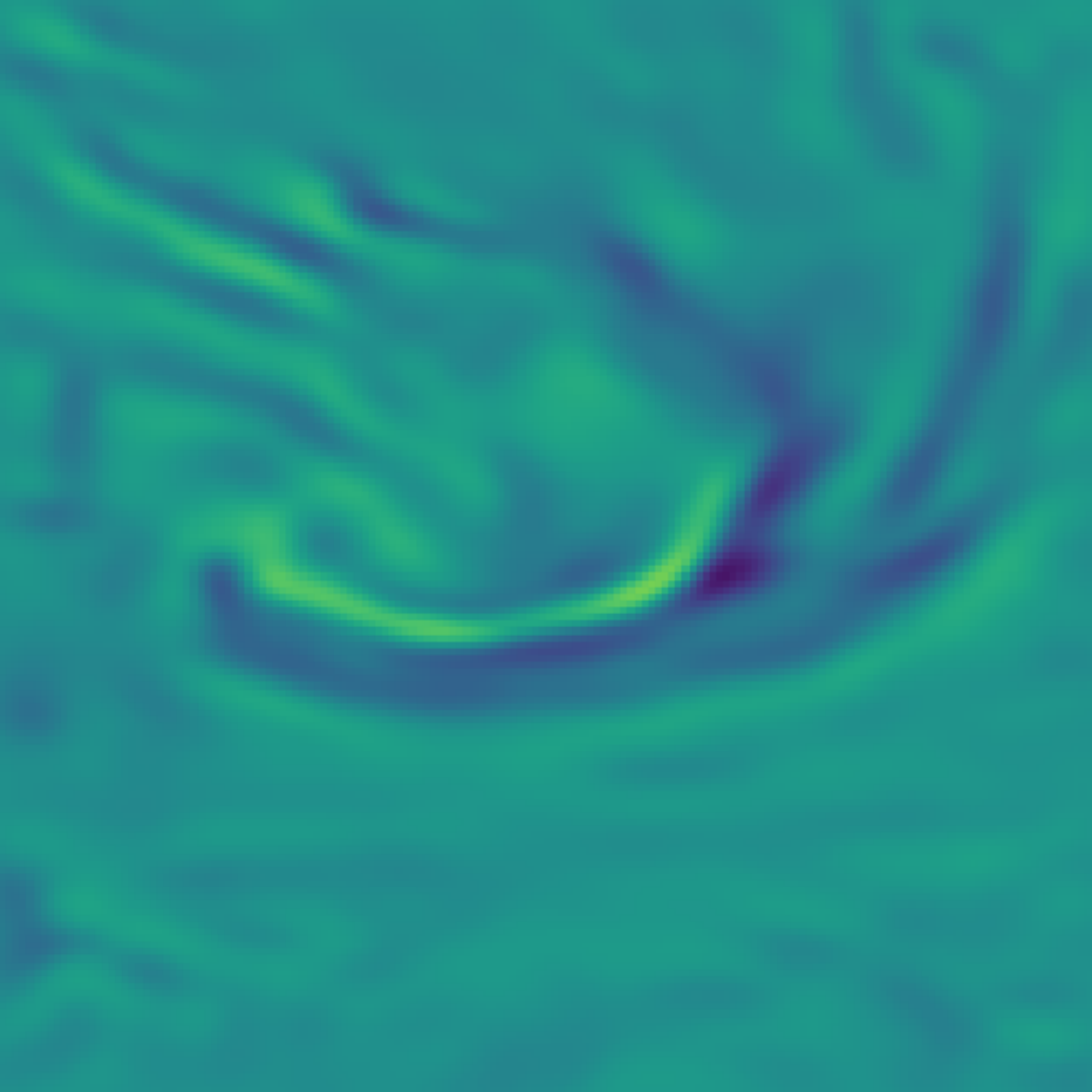}
  \caption{Divergence field (left) and its reconstruction produced by the autoencoder (right). While the reconstructed field lacks finer details, large-scale structures are properly captured by the autoencoder}
  \label{fig:autoencoder_examples}
\end{figure}

Our network used for inpainting follows the same architecture as the autoencoder described above.

When training with either the AtmoDist- or autoencoder-based loss function, we initialize the network with a pretrained version in the same way as we did for the super-resolution already. Furthermore, a small $l2$-loss term is added as a regularization. Specifically, the total loss is given by
\begin{equation}
    L(X_1,X_2) = (1-\alpha) L_\text{content}(X_1,X_2) + \alpha \left\|X_1 - X_2 \right\|_2 ,
\end{equation}
where $\alpha=0.1$ and $L_\text{content}$ is either the already scaled AtmoDist or autoencoder loss. This is done to prevent local minima and artifacts during training.

\subsection{Semivariogram calculation}
\label{sec:appendix:semivariogram}

The semivariogram, given by
\begin{equation}
    \label{eq:semivariogram}
    \gamma(r) = \int (f(x+r) - f(x))^2 \mathrm{d}x
\end{equation}
 can be calculated in different ways.
We approximate the integral that defines it using Monte-Carlo sampling. 
In particular, for each time-step and each lag-distance $r$, 300 random locations and 300 random directions are sampled and the field is evaluated at these points. This procedure is done for the complete evaluation period and in the end the semivariogram is obtained by averaging. 

\begin{figure}
  \centering
  \includegraphics[width=\textwidth]{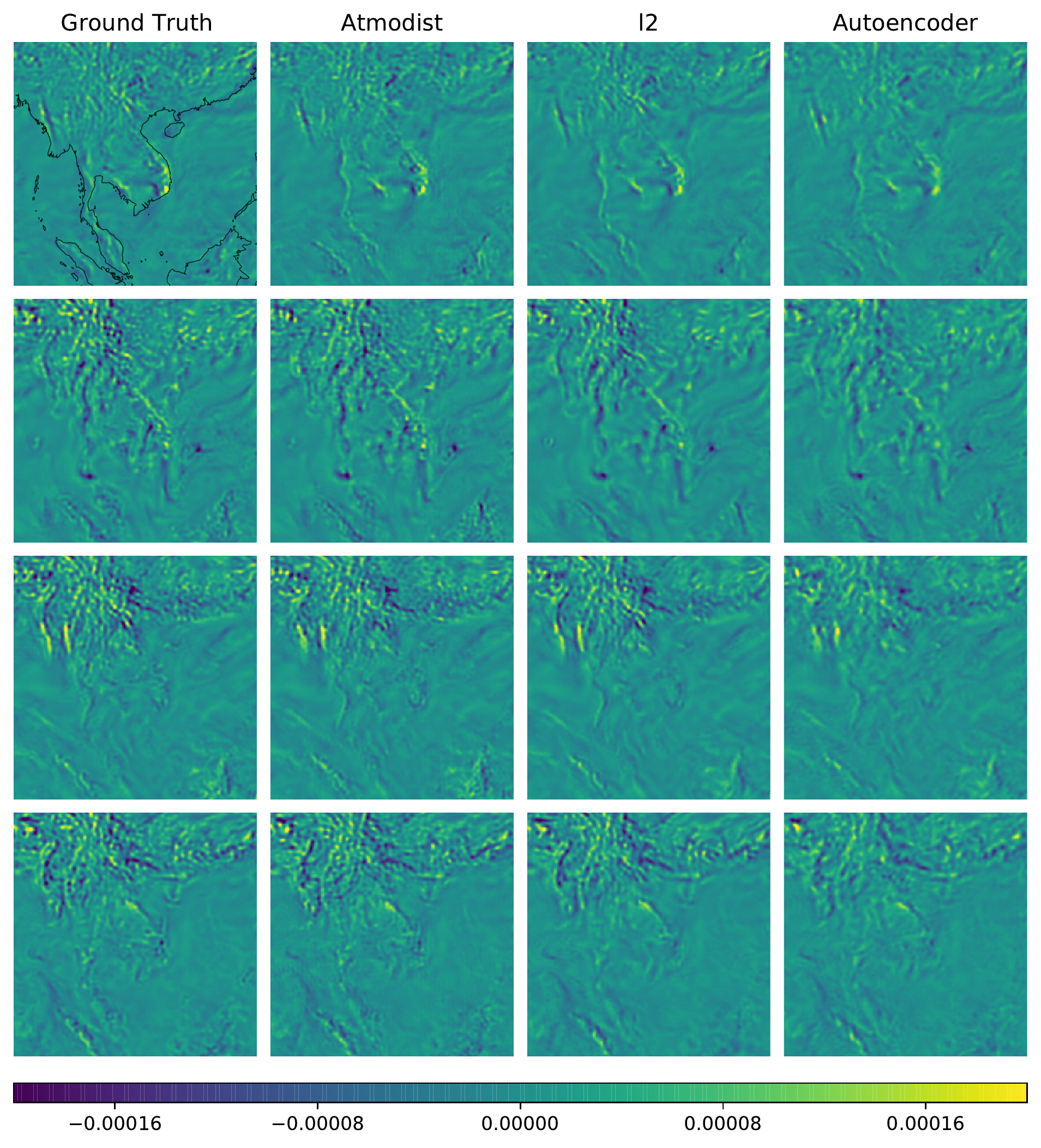}
  \caption{\centering Downscaled divergence fields over the Gulf of Thailand at different timesteps. Coastlines are shown in the first ground truth field and then omitted for better comparability}
  \label{fig:downscaling_examples}
  \label{fig:downscaling_examples1}
\end{figure}

\begin{figure}
  \centering
  \includegraphics[width=\textwidth]{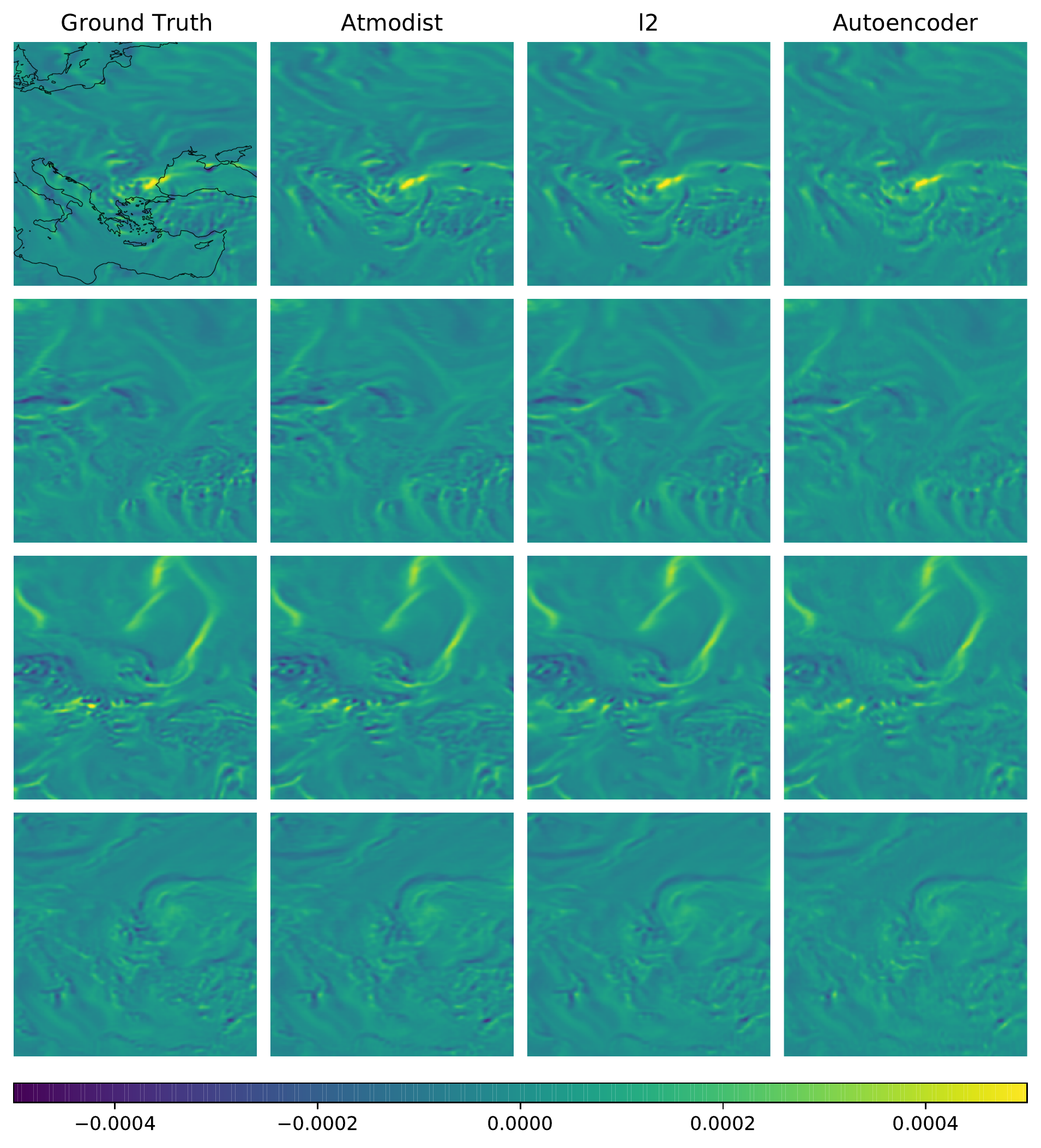}
  \caption{Downscaled vorticity fields over the Mediterranean Sea and Eastern Europe at different timesteps. Coastlines are shown in the first ground truth field and then omitted for better comparability}
  \label{fig:downscaling_examples2}
\end{figure}

\end{appendix}

\end{document}